\documentclass[prd,preprint,superscriptaddress,nofootinbib]{revtex4}

\usepackage{amsmath,amssymb,array,calc,epsfig}
\usepackage{graphicx}

\newlength{\intwidth}

\def\h{\hbox{-}}

\begin{document}

\title{On the Structure of Scattering Amplitudes in ${\cal N}=4$ Super Yang-Mills and ${\cal N}=8$
Supergravity}

\author{Freddy Cachazo}
\email{fcachazo@perimeterinstitute.ca}

\affiliation{Perimeter Institute for Theoretical Physics, Waterloo,
Ontario N2J 2W9, Canada}

\author{David Skinner}
\email{dskinner@perimeterinstitute.ca}

\affiliation{Perimeter Institute for Theoretical Physics, Waterloo,
Ontario N2J 2W9, Canada}

\begin{abstract}

Exploiting singularities in Feynman integrals to get information
about scattering amplitudes has been particularly useful at one-loop
in theories where no triangles or bubbles appear. At higher loops
the integrals possess subtle singularities. In this paper we give
these singularities a physical interpretation and show how they turn
tedious computations into purely pictorial manipulations. We
illustrate our methods with various examples from the computation of
four-particle amplitudes in ${\cal N}=4$ super Yang-Mills and ${\cal
N}=8$ supergravity. Along the way we find clues towards an
understanding {\it i}) of the rung-rule as a consequence of
infra-red singularities, {\it ii}) of the non rung-rule integrals
included in the basis as corrections to the rung-rule and {\it iii})
of the coefficients - including signs - of these two types of
contribution. The role of corrections is to cancel unphysical
singularities generically present in rung-rule integrals. A further
byproduct, coming from the fact such unphysical singularities are
located where conformal cross-ratios become unity, is the
possibility of understanding the dual conformal invariance ansatz
for constructing the basis of four-particle amplitudes in ${\cal
N}=4$ SYM.

\end{abstract}

\maketitle

\section{Introduction}

Scattering amplitudes of massless particles in four dimensions are
fascinating objects with many properties deeply buried in their
Feynman diagram representation. At one-loop, the Passarino-Veltman
reduction procedure~\cite{Passarino:1978jh,van Neerven:1983vr} gives a different representation of the
amplitudes in terms of a simple basis of integrals with coefficients
that encode the details of the particular amplitude under
consideration. A special class of theories possesses an even simpler
basis: the scalar integrals involve only box diagrams. It is known
that ${\cal N}=4$ SYM belongs to this class~\cite{Bern:1994zx} and it has
been hypothesized that so too does ${\cal N}=8$
supergravity~\cite{Bern:1998ug,BjerrumBohr:2006yw}. In this special class of theories,
all one-loop amplitudes can be computed in terms of tree-level
amplitudes using a procedure called quadruple
cuts~\cite{Britto:2004nc}. Quadruple cuts exploit the fact that the
leading singularity of a given scalar integral is unique.

It is tempting to try to apply this procedure to higher loops. In fact, the first
attempt was made some years ago by one of the authors and Buchbinder
in~\cite{Buchbinder:2005wp}. The most obvious difficulty here is that the $L$-loop basis
of integrals is not known for arbitrary number of particles. A more modest problem is that even in the cases where an ansatz for the integral basis exists, the $L$-loop integrals in this basis always contain fewer propagators than the number of integration variables, {\it i.e.} $4L$. In~\cite{Buchbinder:2005wp} the solution to this latter problem was shown to be that, after using four propagators to perform one of the quadruple cut loop integrations, the Jacobian of the change of variables produced one more propagator which could then be used in the next integration. Taking this into account, in this paper we adopt the terminology that the `maximal cut' of an $L$-loop diagram refers to cutting $4L$ propagators - these may be visible in the original diagram, or else arise from hidden singularities\footnote{Our terminology thus differs from that of~\cite{Bern:2007ct}, where the `maximal cut' referred to cutting only the propagators visible in an $L$-loop diagram.}. Once all integration variables are fixed, the coefficient of the integral is given by a sum of products of tree-level amplitudes, where the sum runs over all the possible choices of helicity for the particles running along each internal leg of the diagram. This sum makes the procedure very cumbersome in the case of ${\cal N}=4$ SYM and even more so in ${\cal N}=8$ supergravity, where there are generically many possible internal states to consider. Hence we are faced with two puzzles: why does increased supersymmetry apparently make the computation more difficult rather than simpler? And what is the physical meaning of the new effective propagators (or hidden singularities) of the scalar integrals?

\bigskip

In the first part of the paper, we solve these puzzles. For four-particle amplitudes at any loop order, Ward identities (discussed in~\cite{Bern:1998ug} for both ${\cal N}=4$ SYM and ${\cal N}=8$ supergravity) essentially constrain the above sum of products of tree amplitudes to be a four-particle tree amplitude times a helicity-independent factor, as illustrated in figures~\ref{fig:sym} \&~\ref{fig:sugra}. Applying this observation to the sum of Feynman diagrams with the structure of the diagram under consideration allows us to collapse loop subamplitudes into their corresponding tree-level amplitudes, as we explain in section~\ref{sec:1loop}. The hidden singularities of the scalar integrals depend on the external momenta of these tree subamplitudes, and in section~\ref{sec:higherloops} we show that cutting a particular hidden propagator corresponds to factorizing the tree amplitude in a particular channel. In this way, $L$-loop diagrams are reduced to $(L-1)$-loop diagrams and the process of determining integral coefficients from their maximal cuts is dramatically streamlined.

\bigskip

In section~\ref{sec:physicalsing} we build our familiarity with the new techniques by studying a variety of contributions to four-particle multi-loop amplitudes in planar ${\cal N}=4$ SYM coming from the rung-rule~\cite{Bern:1994zx,Bern:1997nh,Bern:2005iz}. However, the technique is not limited to this case. In section~\ref{sec:gravity} we show that the same methods can be applied with equal ease to the more difficult problem of understanding the coefficients of the basis of multi-loop scalar integrals for ${\cal N}=8$ supergravity. We uncover a restricted notion of `dual conformal invariance'~\cite{Drummond:2006rz,Drummond:2007aua} (reviewed briefly in section~\ref{sec:dci}) that applies to the scalar integrals in ${\cal N}=8$. In studying gravity amplitudes, it is important that we can consider both planar and non-planar contributions to the amplitude, as in gravity there is no real distinction between the two. We show that the methods of sections~\ref{sec:1loop}-\ref{sec:higherloops} apply here also (we expect that non-planar contributions to ${\cal N}=4$ SYM amplitudes can be similarly analyzed, provided the colour-structure is accounted for).

\bigskip

The second part of the paper is devoted to understanding extensions of the technique to more general cases. At present, our method can only be directly applied to a certain class of integrals: those that possess a box with at least two adjacent cubic vertices. However, in section~\ref{sec:lim} we outline arguments that might lead to further generalizations. The main clue comes from the fact that the resummation formulae (figures~\ref{fig:sym} \&~\ref{fig:sugra}) that fuel the whole procedure can also be derived from the infrared singular behavior of the amplitude. Using the infrared singular behavior of an $n$-particle amplitude and some extra assumptions we come close to giving a derivation of the rung-rule.

Of course, it is well known that the rung-rule is incomplete starting at four loops in planar ${\cal N}=4$ SYM~\cite{Bern:2006ew}. In section~\ref{sec:corrections}, we give the first steps towards understanding the non rung-rule integrals. Briefly, certain rung-rule integrals contain unphysical singularities and it is the role of the non rung-rule integrals to `correct' the rung-rule by canceling these singularities. Rung-rule integrals were shown to give rise to dual conformal invariant integrals in~\cite{Bern:2007ct}. It turns out that the location of their unphysical singularities can be expressed in terms of a conformally invariant cross-ratio (originating from the Jacobian of a three-mass scalar box integral). Thus we are guaranteed that the entire basis constructed this way will possess dual conformal invariance. Moreover, as we explain in section~\ref{sec:corrections}, requiring that unphysical singularities are canceled provides a simple rule for determining the relative sign of the non rung-rule integrals, compared to the rung-rule contributions. This sign may be either $\pm1$ depending on the order of the correction.

\section{Preliminaries}
\label{sec:prelims}

Scattering amplitudes of on-shell particles in ${\cal N}=4$ SYM with
gauge group $U(N)$ can be written as a sum over color-stripped
partial amplitudes using the color decomposition (see {\it e.g.}~\cite{Dixon:1996wi}). Each partial
amplitude admits a large $N$ expansion. More explicitly,
\begin{eqnarray}
{\cal A}_n(1,2,\ldots , n) &=& \delta^{(4)}(p_1+p_2+\ldots + p_n)\ {\rm
Tr}(T^{a_1}T^{a_2}\ldots T^{a_n})\ A_n(1,2,\ldots, n) \nonumber\\
&&\qquad\qquad + \ {\rm permutations}\ +\ \ldots
\end{eqnarray}
where the sum is over non-cyclic permutations of the external states (cyclic ones being symmetries of the trace) and the ellipsis represents terms with double and higher numbers of traces. In ${\cal N}=8$ supergravity there is no color-stripping to be done and we write simply
\begin{equation}
{\cal M}_n(1,2,\ldots , n) = \delta^{(4)}(p_1+p_2+\ldots + p_n)\
M_n(1,2,\ldots, n)\ .
\end{equation}
In each case, $A_n$ and $M_n$ may be expanded perturbatively
and we denote the $L$-loop terms by $A_n^{(L)}$ or $M_n^{(L)}$. We
also use $A_n^{(0)}=A_n^{\rm tree}$ and $M_n^{(0)}=M_n^{\rm tree}$.

The (non-vanishing) amplitude that is currently best understood is perhaps the
planar (leading color) four-particle amplitude in ${\cal N}=4$ SYM.
Given that many of our examples and results are related to this, we first review what is known and what has been hypothesized about this amplitude.

\subsection{Rung-rule}
\label{sec:rr}

A remarkably simple rule to generate integrals in the scalar basis
of $A_4^{(L)}$ was proposed some years ago in a series of papers~\cite{Bern:1994zx,Bern:1997nh,Bern:2005iz}. The main motivation was that two-particle cuts can be performed to all loop orders and to all orders in the dimensional regularization parameter. Given any $(L-1)$-loop integral, one produces an element in the $L$-loop basis by adding a propagator (rung) between any two propagators such
that the new integrals does not contain triangles or bubbles. If the
original two propagators carry momentum $\ell_1$ and $\ell_2$, then a
numerator factor of $(\ell_1+\ell_2)^2$ must be added to the new integral. This procedure of adding rungs is known as the rung-rule and turns out to correctly provide the full
basis of integrals up to three loops (in dimensional regularization)~\cite{Bern:2005iz}. The rungs may be added so as to produce diagrams which do not contain any two-particle cuts\footnote{That is, rung-rule diagrams may be two-particle irreducible. The first such example is at four loops.} and the presence of these diagrams obstructs a proof of the validity of the rung-rule. In section~\ref{sec:lim}, we speculate on how this rule can be derived from the IR behavior of one-loop amplitudes.

\subsection{Dual Conformal Integrals}
\label{sec:dci}

In fact, the rung-rule does not give the complete basis of integrals to all-loop orders, as was discovered in~\cite{Bern:2006ew} where the explicit four-loop integrand was found. Two new non rung-rule integrals had to be included. A remarkably simple new ansatz for constructing the complete basis of integrals for the four-particle planar amplitude was then proposed in~\cite{Drummond:2006rz,Drummond:2007aua,Bern:2006ew,Bern:2007ct}.
The proposal is that the complete basis of integrals be given at any loop order by the set of ``dual conformally invariant" integrals. In short, an integral is dual conformally invariant if the integral represented by the dual diagram is conformally invariant. To define the dual diagram, one
assigns a point $x_i$ to each loop and each external region (between
two external legs) of the original diagram. This assignment is unambiguous for planar
diagrams. Each momentum is then given by $p_{ij}=x_i-x_j$ where
$x_i$ and $x_j$ are the points in the two zones separated by the
leg of the original diagram containing momentum $p_{ij}$. The original
momentum-space loop integral may then be rewritten in terms of
integrals over the internal $x$s. These loop integrals are IR divergent and need to be
regularized - since dimensional regularization breaks conformal
invariance, one instead regularizes by taking the momenta of the
external particles to be off-shell. Such a regularization does not necessarily remove all the divergences of an integral, because subdiagrams may still diverge~\cite{Drummond:2007aua}. Only those integrals which are well-defined after this regularization are truly dual conformally invariant, and these constitute the $L$-loop basis.

The notion of dual conformal invariance is undoubtedly a very
powerful one, if still somewhat mysterious at present. There are however a number of limitations to its
applicability. Firstly, it does not extend to ${\cal N}=8$ supergravity
amplitudes, or even non-planar ${\cal N}=4$ SYM amplitudes - at
least not in an obvious way (we will see in section~\ref{sec:gravity} that there is a restricted sense in which it may still apply). Secondly, although assuming dual
conformal invariance fixes the members of the basis of scalar
integrals, it does not fix their relative numerical coefficients.
Finally, the regularization procedure used in~\cite{Drummond:2006rz,Drummond:2007aua} allows for dual conformal invariant diagrams that are not present in dimensional regularization, so that comparison between the different regularization schemes is not immediate~\cite{Nguyen:2007ya}.

\section{One-Loop Amplitudes and Singularities}
\label{sec:1loop}

The use of singularities to constrain the form of amplitudes is
particularly powerful at one-loop in theories with only scalar boxes
in their expansion. The structure of one-loop amplitudes in ${\cal
N}=4$ SYM and in ${\cal N}=8$ supergravity will be of particular interest
in the rest of this paper. In this section we present a small review
which will also serve to set up conventions.

Scalar box integrals are of the form
\begin{equation}
I(K_1,K_2,K_3,K_4) := \int\frac{d^4\ell}{(2\pi)^4}
\frac{1}{\ell^2(\ell-K_1)^2(\ell-K_1-K_2)^2(\ell+K_4)^2}\ .
\end{equation}
Here, the momenta $K_i$ are given as a sum over the momenta of
external particles. The singularities of loop integrals are usually
extracted by `cutting' propagators and replacing them with delta
functions, {\it i.e.} by removing the principal value part, for example
$1/((\ell-K_1)^2+i\epsilon) \to
\delta^{(+)}\left((\ell-K_1)^2\right)$. When all four propagators
are cut \cite{Britto:2004nc}, there are no solutions for $\ell$ in
real (Minkowskian) momentum space. One way to avoid this difficulty, as
was done in \cite{Britto:2004nc}, is to work in spacetime of
signature $(++--)$. However, in the context of studying the analytic
properties of amplitudes, it is perhaps more natural to consider
analytically continuing the integrand into complexified momentum
space $\Bbb{C}^4$. We can view the usual integrals this way, where
the contour is taken to be a copy of real momentum space inside
$\Bbb{C}^4$ (and is ill-defined due to infra-red divergences). The
residue of the leading singularity is obtained by instead taking the
contour to be a real $T^4\subset\Bbb{C}^4$ which encircles the poles
from the propagators\footnote{This is in the same spirit as the manipulations done in a different context in~\cite{Vergu:2006np} in order to study factorization limits in the connected prescription for Yang-Mills amplitudes~\cite{Roiban:2004yf}.}. That is, we will consider integrals of the
type
\begin{equation}
\label{eq:contourint}
\Delta I :=\oint_\Gamma \frac{d^4\ell}{(2\pi{\rm i})^4}\
\frac{1}{\ell^2(\ell-K_1)^2(\ell+K_2)^2(\ell+K_2+K_3)^2}\ ,
\end{equation}
where $\Gamma = \left\{\ell : |f_i(\ell)|=\epsilon_i\right\}$, with $f^{-1}_i(\ell)$ being a propagator and $\epsilon_i$ some small positive number\footnote{The overall orientation of the contour will not concern us, because we will always be comparing the result of imposing a choice of contour on both the sum of Feynman diagrams and on the members of the basis of scalar integrals. Reversing the contour orientation would simply lead to a change in sign of both sides.}. Note that each propagator becomes singular on a
quadric hypersurface of codimension 1 in $\Bbb{C}^4$, so that there
is just enough room to encirle it with an $S^1$ factor of $\Gamma$.

For generic external momenta, the four quadrics will intersect in isolated points ({\it i.e.} cutting all four propagators fixes the loop momentum, up to a finite set) so that the integral is
well-defined without the use of dimensional regularization or virtuality of external states. The residue of such a contour integral is just the Jacobian of the change of variables from $\ell_\mu$ to
$f_i$ (see {\it e.g.}~\cite{GH}):
\begin{equation}
\Delta I = \sum_{\ell: f_i(\ell)=0}\det\left(\frac{\partial
f_i}{\partial\ell_\mu}\right)^{-1} \ .
\end{equation}
Note that this is the same result one would obtain from replacing
the propagators by delta functions in $(++--)$ signature.

We can equally well write the integral in terms of the dual
variables $x_i$ of~\cite{Drummond:2006rz}. The contour is again
specified by a product of $S^1$ factors of the form
$|x_{ij}^2|=\epsilon$. As is well known, provided it does not cross
any singularities, the contour may be deformed arbitrarily without
affecting the value of the integral\footnote{The contour integral is
really a pairing between the homology class of the contour in
$\Bbb{C}^4-X_{\rm sing}$ where $X_{\rm sing}$ is the union of
quadrics upon which each propagator becomes singular, and the {\v
C}ech cohomology class represented by the integrand.} and hence the
integral has the same transformation properties (in particular under
the conformal group) as the integrand. In particular, if the integrand of
equation~({\ref{eq:contourint}) is multiplied by $st$, where $s=(K_1+K_2)^2$ and
$t=(K_2+K_3)^2$, then both the integrand and contour integral
are invariant under dual conformal transformations. Thus, by computing the loop
integral over a $T^4$ rather than $\Bbb{R}^4$ contour, (dual)
conformal covariance is preserved.

\subsection{One-Loop Amplitudes in ${\cal N}=4$ SYM and ${\cal N}=8$ Supergravity}
\label{sec:symsupergravity1loop}

At one-loop in ${\cal N}=4$ SYM and in ${\cal N}=8$ supergravity (provided
the no-triangle hypothesis holds), the technique of replacing a
divergent integral $I$ by the maximally cut ({\it i.e.} contour) integral $\Delta I$ works
well~\cite{Britto:2004nc,Bern:2005bb}. Any one-loop amplitude may be expressed both as a sum over Feynman diagrams, and also in terms of scalar box integrals\footnote{For a superspace effective action perspective on one-loop amplitudes, see~\cite{Kallosh:2007ym}.}:
\begin{equation}
A_n^{(1)}= \sum\left\{\hbox{1-loop Feynman diagrams}\right\}
= \sum_{\cal I} B_{\cal I}\times I(K_1^{\cal I},K_2^{\cal
I},K_3^{\cal I},K_4^{\cal I})
\label{eq:quadruple}
\end{equation}
where the second sum is over all partitions ${\cal I}$ of
$\{1,2,\ldots, n\}$ into four non-empty sets, $K_i^{\cal I}$ equals
the sum of the momenta in the $i^{\rm th}$ subset of partition
${\cal I}$ and the $B_I$ are coefficients to be determined. In the case of ${\cal
N}=4$ SYM one considers only the Feynman diagrams and partitions that respect the color
ordering.

\begin{figure}
\includegraphics[scale=0.50]{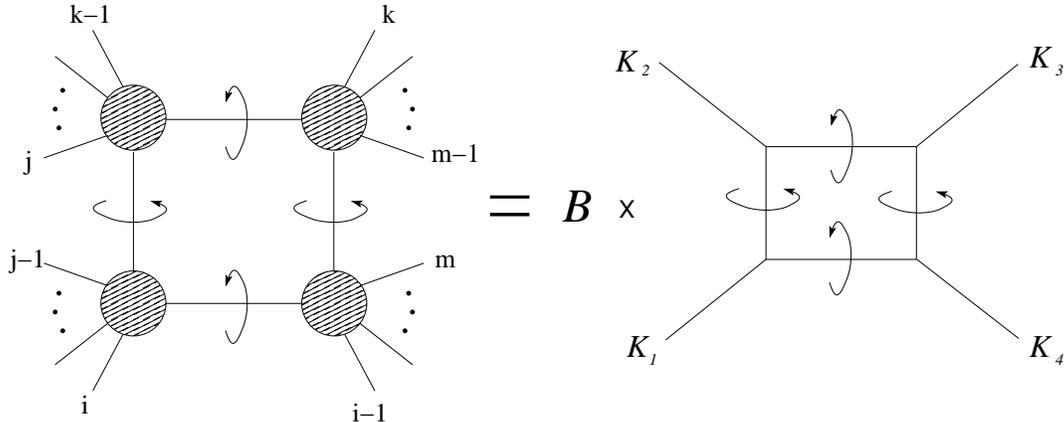}
\caption{Computation of a coefficient using the leading singularity
of a box. The lines circling the propagators represent the $T^4$ contour of integration. The left hand side of the figure represents the sum of all 1-loop Feynman diagrams - note that only those Feynman diagrams that contain the displayed propagators actually contribute to this particular contour integral.}
\label{fig:quadruple}
\end{figure}

The computation of the amplitude is thus reduced to finding the
coefficients $B_{\cal I}$. To do so, we analytically continue (the integrands of) both the Feynman diagrams and scalar box $I(K_1,K_2,K_3,K_4)$ in equation~(\ref{eq:quadruple}) and consider a generic partition
$$
{\cal J}=\{ i,\ldots, j-1; j\ldots, k-1;k,\ldots,m-1;m,\ldots, i-1\}
$$
as shown in figure~\ref{fig:quadruple}. The right hand side of this figure represents a scalar box integral, integrated over the $T^4$ contour discussed above. The left hand side of the figure is supposed to represent a sum over all 1-loop Feynman diagrams, integrated over the same contour. With the choice of contour corresponding to the scalar diagram on the right, the only contribution comes from Feynman diagrams with the indicated structure - {\it i.e.} those that contain the appropriate propagators. For the scalar box, the contour integral simply gives the coefficient $B_{\cal J}$ times the Jacobian $\det\left(\partial f_i/\partial\ell_\mu\right)^{-1}$, summed over the number of solutions to the equations $f_i(\ell)=0$. The contour integral on the Feynman diagram side involves precisely the same Jacobian, but now the Feynman diagrams must also be taken into account when computing the residue. This is easy to do: by the standard LSZ argument, the limit of these Feynman diagrams as the specified propagators go on-shell is simply the product of the corresponding (colour-stripped) amplitudes, summed over all possible helicity configurations for the internal particles. Equating both sides we obtain, for each complex solution $\ell_*$ of $f_i(\ell)=0$
\begin{equation}
\left.B_{\cal J}\times \det\left(\frac{\partial f_i}{\partial\ell_\mu}\right)^{-1}\right|_{\ell=\ell_*}
= \left.\det\left(\frac{\partial f_i}{\partial\ell_\mu}\right)^{-1}\sum_{\rm Multiplet}\prod_{i=1}^4
A^{{\rm tree}\ (i)}\right|_{\ell=\ell_*}\ ,
\label{eq:oneloopsum}
\end{equation}
where the sum is over all the members of the multiplet running along each of the internal lines and $A^{{\rm tree}\ (i)}$ are the tree-level amplitudes located at each vertex of the box in the partition ${\cal J}$, as shown in figure~\ref{fig:quadruple}. Note that we are free to view each internal line as representing the full supermultiplet: if, for a certain choice of external states, selection rules forbid some members of the multiplet from appearing in propagators of the original 1-loop diagram, this will be accounted for by the vanishing of the corresponding partial amplitude in equation~(\ref{eq:oneloopsum}).

\bigskip

In the particular case of four-particle colour-ordered amplitudes in ${\cal N}=4$ SYM, there is only one partition to consider. There are two complex solutions to $f_i(\ell)=0$: if the propagators are labelled as in equation~(\ref{eq:contourint}) these are given by $\ell_*=\{\alpha\lambda^{(1)}\widetilde\lambda^{(2)},\ \widetilde\alpha\lambda^{(2)}\widetilde\lambda^{(1)}\}$ with $\alpha=[14]/[24]$ and $\widetilde\alpha=\langle14\rangle/\langle24\rangle$. When evaluated at either solution, the Jacobian is simply $1/(st)$. The sum over products of tree amplitudes may be explicitly performed and for external gluons one obtains $\sum_{\rm Multiplet}\prod_{i=1}^4 A_3^{{\rm tree}\, (i)} = st\times A_4^{\rm tree}(k_1,k_2,k_3,k_4)$, again at each solution $\ell_*$. Consequently the coefficient of the four-particle box integral in ${\cal N}=4$ SYM is fixed to be
\begin{equation}
\label{eq:SYMprod}
B=st\ A_4^{\rm tree}(k_1,k_2,k_3,k_4)
\end{equation}
when the external particles are gluons. Using Ward
identities, four-particle amplitudes where the
external states are any members of the ${\cal N}=4$ multiplet may be
related to the amplitudes with external gluons~\cite{Bern:1998ug}; the only difference
is a helicity-dependent factor that is the same to all loop orders.
In particular, the ratio $A^{(1)}_4/A^{\rm tree}_4$ is independent of
helicity (whenever $A^{\rm tree}_4\neq0$). Consequently, the Ward identities assure us that the coefficient of a scalar box for a four-particle 1-loop amplitude is given by equation~({\ref{eq:SYMprod}), even when the external states are arbitrary members of the ${\cal N}=4$ multiplet. This observation will be a key simplifying idea for the rest of the paper, and is summarized in figure~\ref{fig:sym}. Once again, the left hand side of this figure represents the sum of all 1-loop Feynman diagrams (with a particular colour ordering), analytically continued to complexified momentum space and integrated over the indicated contour. The residue of this contour integral is just the four-particle tree amplitude.

\begin{figure}
\includegraphics[scale=0.45]{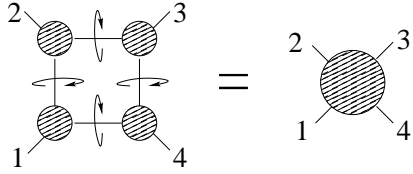}
\caption{Main resummation formula in ${\cal N}=4$ SYM. The left of the diagram represents the sum of all 1-loop Feynman diagrams, integrated over the displayed contour. The external states may be any members of the ${\cal N}=4$ multiplet.}
\label{fig:sym}
\end{figure}

\bigskip

In the case of ${\cal N}=8$ supergravity there is no notion of cyclic ordering of the external states, so for four-particle 1-loop amplitudes we have\footnote{We will write $B$ for the coefficients of scalar integrals in ${\cal N}=4$ SYM and $C$ for the coefficients in ${\cal N}=8$ supergravity.}
\begin{eqnarray}
\label{eq:grav}
M_4^{(1)}(k_1,k_2,k_3,k_4)
&=&\sum \left\{\hbox{1-loop Feynman diagrams}\right\}\\
&=&C_{1234} I(k_1,k_2,k_3,k_4) + C_{1243}I(k_1,k_2,k_4,k_3) +
C_{1324}I(k_1,k_3,k_2,k_4)\nonumber
\end{eqnarray}
in terms of scalar boxes. Choosing a $T^4$ contour to isolate one of
the scalar box integrals, say the first, one again obtains at each solution
\begin{equation}
\frac{C_{1234}}{st} = \frac{1}{st}\sum_{\rm Multiplet}\prod_{i=1}^4 M_3^{{\rm tree}\, (i)}\ ,
\end{equation}
but now for gravity the sum of products of amplitudes gives
\begin{equation}
\sum_{\rm Multiplet}\prod_{i=1}^3 M_3^{{\rm tree}\, (i)}= stu\times M_4^{\rm tree}(k_1,k_2,k_3,k_4)\ ,
\label{eq:gravityproduct}
\end{equation}
so that $C_{1234}=stu\ M^{\rm tree}$. Note that the extra factor of ({\it momentum})$^2$ here, compared to Yang-Mills, is in agreement with simple dimensional analysis. Equation~(\ref{eq:gravityproduct}) may be derived directly, by summing over the entire ${\cal N}=8$ multiplet on the left hand side, or else more simply by use of a linear combination of the gravitational infra-red relations given in~\cite{Dunbar:1995ed}. We will make further use of these IR relations in section~\ref{sec:ir}. Repeating the calculation with different choices of contours (or taking a different linear combination of the IR relations) yields $C_{1234}=C_{1243}=C_{1324}$, although $s$, $t$ and $u$ are permuted in the intermediate steps. Once again, Ward identities show that the ratio $M^{(1)}_4/M^{\rm tree}_4$ is helicity-independent. Consequently, the contour integral over the sum of 1-loop Feynman diagrams must always give either\footnote{Depending only on the ordering of the momentum labels - we will say more about this in section~\ref{sec:gravity}.} $s$, $t$ or $u$ times $M^{\rm tree}_4$, no matter which external supergravity states are considered. This relation is depicted in figure~\ref{fig:sugra}.

\begin{figure}
\includegraphics[scale=0.45]{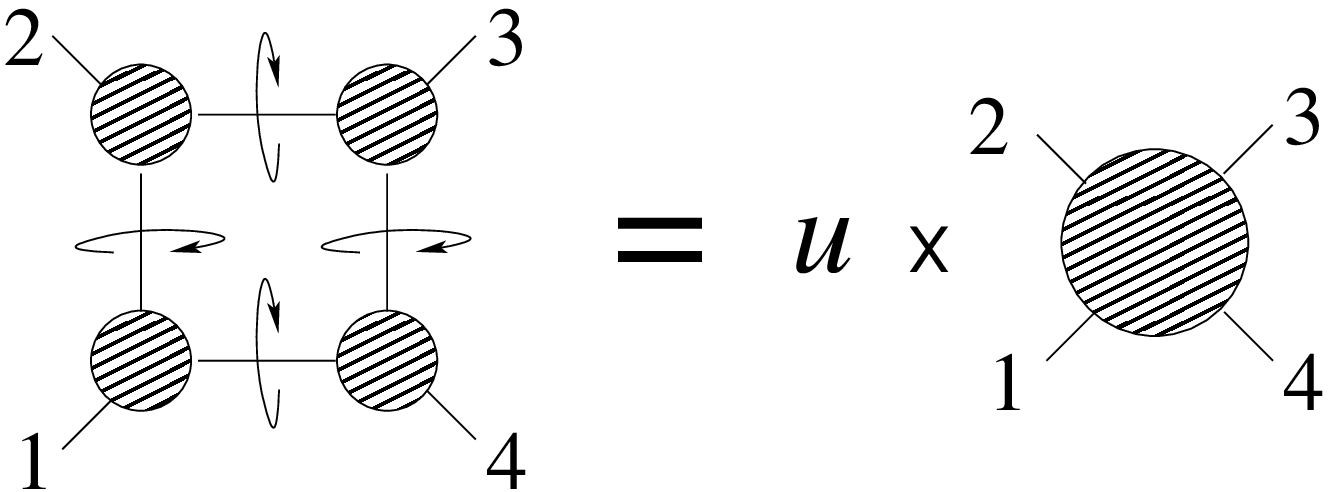}
\caption{Main resummation formula in ${\cal N}=8$ supergravity. The external states may be any members of the ${\cal N}=8$ supermultiplet.}
\label{fig:sugra}
\end{figure}

\section{Generalization to Higher Loops}
\label{sec:higherloops}

In generalizing the above leading singularity technique to higher
loops, one encounters various obstacles. The most serious of these
is that the basis of scalar integrals in terms of which the $L$-loop
amplitude should be expanded is not known in general. (As mentioned
earlier, for four-particle planar amplitudes in ${\cal N}=4$ SYM,
the basis is conjectured to be given by the set of dual
conformally invariant
integrals~\cite{Drummond:2006rz,Drummond:2007aua,Bern:2006ew,Bern:2007ct}.)
In ${\cal N}=8$ supergravity or non-planar ${\cal N}=4$ SYM, the
basis is known to three loops~\cite{Bern:2007hh}.}. We will discuss this in section~\ref{sec:lim}, although we do not yet have a complete understanding.

A related issue is that, while the leading singularity completely specifies the one-loop integral, it
is not clear whether leading singularities can isolate particular members of the $L$-loop basis: different four- and five-loop integrals can become linearly dependent when evaluated on lower-order singularities~\cite{Bern:2007ct}. These relations would seem to indicate that in general, there is no way to distinguish all the different members of an $L$-loop basis merely by looking at their leading singularities. However, we note that the relations discovered in~\cite{Bern:2007ct} always involve at least one integral that is not truly a member of the basis: its integrand is dual conformally invariant, but the integral is not well-defined because of divergent subloops~\cite{Drummond:2007aua}. It seems reasonable to exclude such ill-defined integrals {\it a priori}, in which case there are no known ambiguities in the leading singularities of the integrals.

\bigskip

A second difficulty is that an $L$-loop diagram (with $L>1$)
contains fewer than $4L$ propagators. In particular, four-particle
diagrams with only trivalent vertices involve $3L+1$ propagators and
hence encircling each propagator with an $S^1$ factor does not
completely fix the contour. A way around this obstacle was proposed
in~\cite{Buchbinder:2005wp} where it was observed that if searches
for singularities one loop at a time, {\it i.e.} one picks a $T^4$
contour for a given loop variable, then the resulting Jacobian has
new singularities as a function of the remaining loop variables. We
call these singularities of the Jacobian `hidden singularities' as
opposed to the `visible singularities' of the explicit
propagators. To completely specify the integral, we need to pick a $4L$-dimensional contour. We will say that a contour of topology $T^{4L}$ encircling both the visible and hidden singularities is a maximal cut\footnote{Note that what is called the maximal cut
in~\cite{Bern:2007ct} corresponds to encircling only the visible
singularities.}.

Let us illustrate this with the simplest $L$-loop example: the
$L$-loop planar ladder shown in figure~\ref{fig:ladder}. There are
$4L$ integration variables but only $3L+1$ propagators and hence
only $3L+1$ visible singularities. If we choose the contour
\begin{equation}
\Gamma_p:= \left\{|p^2|=\epsilon,|(p+k_3)^2|=\epsilon,|(p-k_4)^2|=\epsilon,
|(p+\ell+k_3)^2|=\epsilon\ :\ p\in\Bbb{C}^4\right\}
\end{equation}
then the integral over $p$ becomes
\begin{equation}
\oint_{\Gamma_p} \frac{d^4p}{(2\pi{\rm i})^4} \frac{1}{p^2(p+k_3)^2(p-k_4)^2(p+\ell+k_3)^2} =
\frac{1}{(k_3+k_4)^2(\ell+k_3)^2}\ .
\end{equation}
Note that the Jacobian contains a new propagator for $\ell$. Now
defining a contour $\Gamma_{\ell}$ to encircle the singularities of
the $\ell$ propagators - including the new one from the Jacobian -
we likewise find that the $\ell$ integral produces
$1/((k_3+k_4)^2(q+k_3)^2)$, yielding a $q$-dependent propagator.
Hence the process can be iterated until all $4L$ integrals have been
performed. By induction, the result of this $T^{4L}$ contour integral
over the $L$-loop ladder of scalar propagators is just
\begin{equation}
\oint_{\Gamma}\hbox{$L$-loop scalar ladder} = \frac{1}{s^Lt}
\label{eq:ladderJac}
\end{equation}
if the ladder is along the $s$-channel.

\begin{figure}[t]
\includegraphics[scale=0.45]{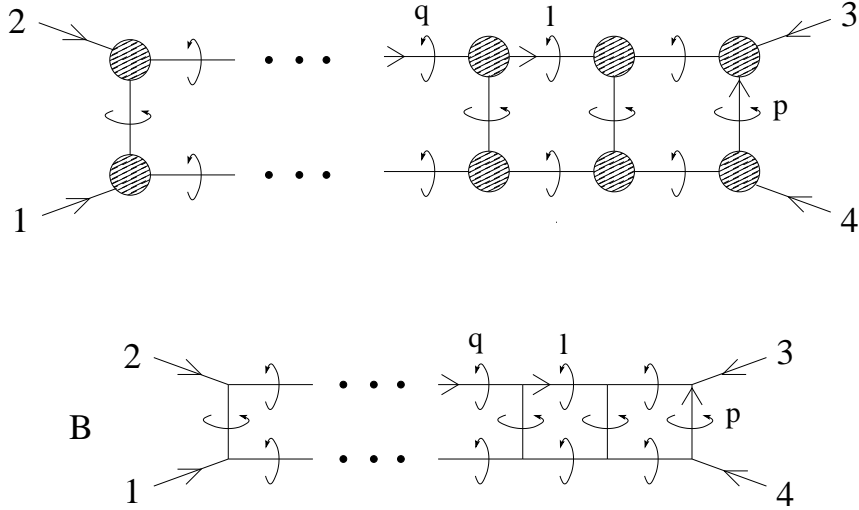}
\caption{The $L$-loop ladder integral, shown for both the scalar integrals and as a factorization channel of the $L$-loop Feynman diagrams.} \label{fig:ladder}
\end{figure}

\subsection{Physical Interpretation of the Hidden Singularities}
\label{sec:physicalsing}

We now consider applying this procedure to the sum of $L$-loop
Feynman diagrams, and for the moment restrict our attention to ${\cal N}=4$ SYM.
Only those diagrams that share the same (visible) propagators as the ladder integral
under consideration will contribute to the contour integral in figure~\ref{fig:ladder}. Suppose, as
in~\cite{Buchbinder:2005wp}, we first perform the integral over the
$T^{3L+1}$ encircling the visible propagators. As at one-loop, the
residues of the Feynman diagrams that contribute in this contour are
given by a sum of products of $2L+2$ three-particle amplitudes, as
shown in figure~\ref{fig:ladder}. Here, the sum is over all members of the ${\cal
N}=4$ SYM multiplet that can run in each internal leg. This sum of products of
amplitudes was found to be independent of the remaining integrated momenta at two
loops in~\cite{Buchbinder:2005wp} and up to five loops in~\cite{Bern:2007ct}.
Assuming this remains true for $L$ loops, the ladder coefficient is fixed to be
\begin{equation}
\label{eq:sumprod}
B^{L\h{\rm loop}\ {\rm ladder}} = \sum_{\rm Multiplets} \prod_{i=1}^{2L+2}A^{{\rm tree}(i)}_3\ .
\end{equation}
However, it is no longer practical to evaluate the right hand side
of~\eqref{eq:sumprod} analytically; the number of terms in the sum
proliferates so rapidly with increasing $L$ that the calculation
soon has to be done numerically.

As noted above, performing only this `visible cut' does not fix the
integrals completely. A very natural question which was not
addressed in~\cite{Buchbinder:2005wp} is the physical meaning of the
remaining, hidden singularities. The answer to this question is
simple. Recall that, in the case of ${\cal N}=4$ SYM, the residue of all four-particle one-loop amplitudes is equal to the corresponding tree amplitude. We thus use figure~\ref{fig:sym} to replace the four three-particle amplitudes that arise when performing the $\Gamma_p$ contour by the four-particle
amplitude. This four-particle amplitude has two factorization channels,
corresponding exactly to the two poles we encountered before in the Jacobian
from integrating out $p$, namely $\left((k_3+k_4)^2(\ell+k_3)^2\right)^{-1}$.
Thus, taking the residue of the four-particle amplitude as $(\ell+k_3)^2\to0$
corresponds to factorizing it in the $(\ell+k_3)^2$ channel and taking
the product of the two resulting three-particle amplitudes, as shown
in figure~\ref{fig:ladder2}.

\begin{figure}[t]
\includegraphics[scale=0.40]{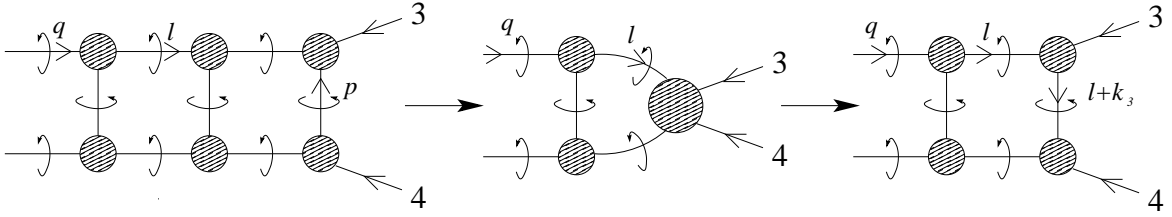}
\caption{The Jacobian provides factorization channels of the $(L-1)$-loop amplitude. In a ladder integral, one of these factorization channels does not contribute because for generic, fixed external momenta, $s$ is always non-zero.}
\label{fig:ladder2}
\end{figure}

So the Feynman diagrams with the structure of an
$L$-loop ladder are reduced by this procedure to Feynman diagrams
with the structure of an $(L-1)$-loop ladder. Just as for the scalar integrals, the
process may be iterated. The final 1-loop box integral is evaluated once again using
figure~\ref{fig:sym} so that we find
\begin{equation}
\oint_{\Gamma_{\rm ladder}}
\hspace{-0.5cm}\left(\hbox{Sum of $L$-loop colour-ordered Feynman diagrams}\right)\
=\  A_4^{\rm tree}(k_1,k_2,k_3,k_4)
\end{equation}
where the contour is the same as in equation~(\ref{eq:ladderJac}). Overall, we
have found that the coefficient of a four-particle, $L$-loop, $s$-channel scalar
ladder integral for ${\cal N}=4$ SYM is
\begin{equation}
B^{L\h{\rm loop}\ {\rm ladder}}=s^L t\times A_4^{\rm tree}(k_1,k_2,k_3,k_4)
\label{eq:LladderSYM}
\end{equation}
as is well known.

Note that it is important that the reduction of four three-particle
amplitudes to the four-particle tree amplitude be valid for arbitrary
external states in the supermultiplet: the external legs of the
first loop can be internal legs of the overall diagram and
these involve a sum over all helicities. Supersymmetric Ward
identities imply that the process of reduction from $L$-loop to
$(L-1)$-loop diagrams is helicity-independent so that all cases can be considered simultaneously.
This makes the computation simple enough to be carried out entirely pictorially;
the sum over helicities for the `external' legs of our initial loop that are really internal legs of the full diagram is automatically incorporated in figure~\ref{fig:ladder2}.

\bigskip

In the above discussion of ladder diagrams, the $1/s$ factor in each Jacobian
played only a passive role. Since this factor is
independent of all loop momenta, the integration contour cannot be chosen so as to exhibit the
collinear singularity when $s\to 0$. However, in more general rung-rule
diagrams, both factors in the Jacobian may depend on momenta from
other loops. In this more general case, it is necessary to check that the scalar diagram has the correct properties in each of the possible factorization limits.

\begin{figure}[t]
\includegraphics[scale=0.40]{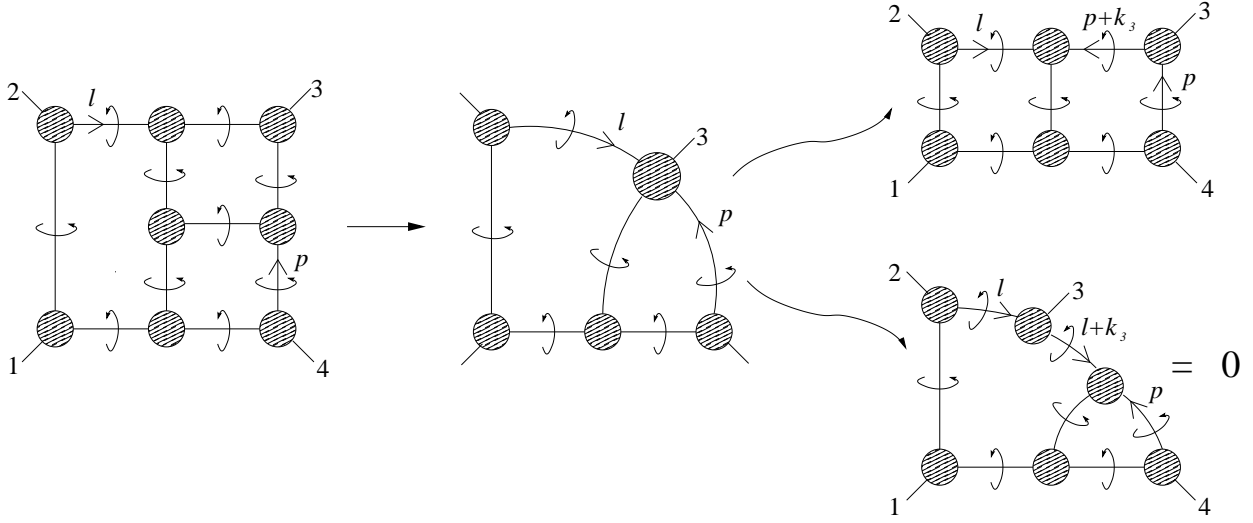}
\caption{Reduction of the tennis court diagram to a 2-loop ladder. The scalar integral coefficient must involve a factor of $(\ell+k_3)^2$ to prevent it having a contribution in a factorization channel that leads to Feynman diagrams containing triangles.}
\label{fig:tennis}
\end{figure}

Consider for example figure~\ref{fig:tennis} which shows the set of
Feynman diagrams with the structure of a three-loop tennis court
diagram. In the first step we integrate out the upper right box;
for scalar diagrams the resulting Jacobian $1/((\ell+k_3)^2(p+k_3)^2)$ allows us to
continue by choosing a contour encircling either of these two `hidden' propagators. The resulting contours would lead to factorizations of the sum of Feynman diagrams into one of the two diagrams shown on the right of the figure. The upper diagram is simply the two-loop ladder diagram; this may be replaced by the two-loop ladder scalar diagram with its coefficient from equation~(\ref{eq:LladderSYM}). The lower diagram contains a one-loop triangle subdiagram\footnote{Note that although the blobs in figure~\ref{fig:tennis} represent a sum of all (colour-ordered) Feynman tree diagrams with the given external legs, blobs with only three external legs cannot hide any propagators. Thus the diagram on the lower right of figure~\ref{fig:tennis} really does contain triangles.}. In the class of theories under consideration such Feynman diagrams sum to zero. Consequently we must have
\begin{eqnarray}
{\rm Res}_{\ (p+k_3)^2\to0}\ \frac{B^{\rm tennis}}{(p+k_3)^2(\ell+k_3)^2} &=& B^{2\h{\rm loop}\ {\rm ladder}}\\
{\rm Res}_{\ (\ell+k_3)^2\to0}\ \frac{B^{\rm tennis}}{(p+k_3)^2(\ell+k_3)^2} &=& 0
\label{eq:tennisres}
\end{eqnarray}
and hence we take
\begin{equation}
B^{\rm tennis} = (\ell+k_3)^2 B^{2\h{\rm loop}\ {\rm ladder}} = s^2t(\ell+k_3)^2 A_4^{\rm tree}\ ,
\end{equation}
in agreement with the rung-rule. We repeat that the manipulations performed in the figure are valid when the external states are arbitrary members of the supermultiplet. The sum over helicities in the internal lines is accounted for automatically.

Sometimes it may occur that both possible factorization channels lead to diagrams that are already part of the $(L-1)$-loop basis. For example, consider the four-loop rung-rule diagram shown in figure~\ref{fig:4loop} (the first rung-rule diagram that does not contain any two-particle cuts). Integrating out the upper left box and considering factorization limits gives
\begin{eqnarray}
{\rm Res}_{\ (r+k_2)^2\to0}\ \frac{B^{4\h{\rm loop}}}{(r+q)^2(r+k_2)^2} &=& B^{\rm tennis}\\
{\rm Res}_{\ (r+q)^2\to0}\ \frac{B^{4\h{\rm loop}}}{(r+q)^2(r+k_2)^2} &=& \widetilde{B}^{\rm tennis}\ ,
\label{eq:4loopres}
\end{eqnarray}
where $\widetilde B^{\rm tennis}$ simply means the coefficient of the tennis diagram at the bottom right of figure~\ref{fig:4loop}, {\it i.e.} with $s\leftrightarrow t$. These requirements are likewise satisfied by
taking
\begin{eqnarray}
B^{4\h{\rm loop}} &=& (r+q)^2 B^{\rm tennis} + (k_2+r)^2 {\widetilde B}^{\rm tennis} \\
&=& s^2t(r+q)^2(r+k_{23})^2A_4^{\rm tree}(1,2,3,4)
+ st^2(k_2+r)^2(r+q+k_1)^2A_4^{\rm tree}(1,2,3,4)\nonumber
\end{eqnarray}
where $k_{23}:=k_2+k_3$ and the other momentum factors correspond to the labelling of figure~\ref{fig:4loop}. Each of the two possible rung-rule numerators associated with a scalar diagram of this topology are thus accounted for. There is a single other 4-loop diagram with no two-particle cuts, obtained from this one by exchanging $k_1\leftrightarrow k_2$, $k_3\leftrightarrow k_4$. It can be analyzed similarly.

\begin{figure}[t]
\includegraphics[scale=0.40]{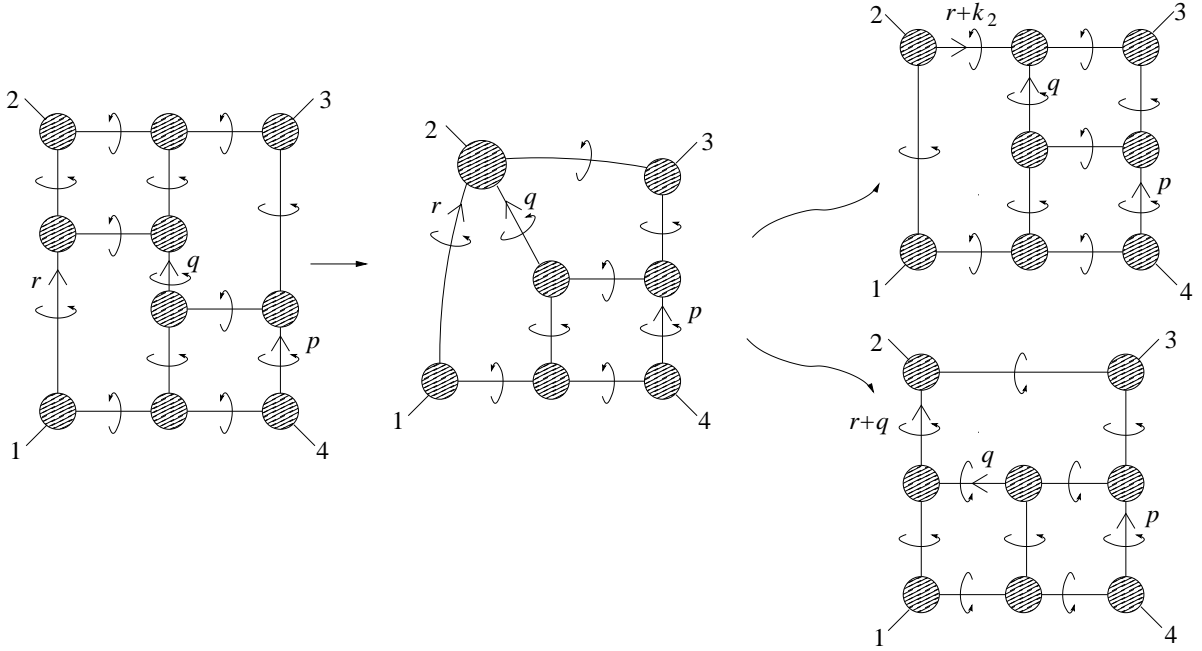}
\caption{Reduction of a 4-loop diagram to tennis court diagrams.}
\label{fig:4loop}
\end{figure}

\bigskip

To summarize, all rung-rule numerators can be understood as ensuring that the scalar
integrals have the same residues as the corresponding Feynman diagrams in all possible factorization channels. In the simplest case, one factorization channel may lead to triangle diagrams, so in any theory obeying the no-triangle hypothesis, numerator factors must be present to cancel singularities in the scalar integrals that would otherwise lead to unphysical contributions in these channels. The overall factors of $s$ and $t$ instead prevent factorization along a one-particle reducible channel, but can be thought of on the same footing by imagining a given diagram as a subdiagram of some larger integral. Equivalently, one can view these factors as taking care of the collinear behaviour of the loop diagram.

In general, we expect our technique to explain the coefficient of all rung-rule diagrams up to seven loops. We have checked explicitly up to five loops that all factorization channels that lead to triangle subdiagrams are prevented by rung-rule numerators. At eight loops there is a rung-rule diagram that does not contain any boxes; we expect that this diagram may be handled similarly, except that one must begin by choosing any four propagators from a common pentagon, and evaluating the remaining propagator as part of the residue. Starting at four loops, the scalar basis is known to involve non rung-rule diagrams. We comment further on these in section~\ref{sec:lim}.

\section{${\cal N}=8$ Supergravity and Non-planar diagrams}
\label{sec:gravity}

It should be clear that much of the above discussion goes through unchanged in ${\cal N}=8$ supergravity, provided it obeys the no-triangle hypothesis~\cite{Bern:1998ug,BjerrumBohr:2006yw}. Thus, just as in SYM, the scalar integrals must come with coefficients to ensure they have the same contribution as the Feynman diagrams in any factorization channel - in particular so they vanish on contours leading to triangle subdiagrams. There will also be factors of kinematic invariants from the overall Jacobian whose role is both to ensure the scalar integrals have the correct collinear limits, and to give the correct factorization limits when the rung-rule diagram appears inside a larger diagram.

The difference comes from the contour integral over the Feynman diagrams. In ${\cal N}=8$ supergravity, we must use equation~(\ref{eq:gravityproduct}) to replace the three-particle amplitudes by the four-particle tree amplitude, times the Mandelstam $u$ of the particular subloop that is being integrated out. Note that supergravity tree amplitudes have singularities in each of their $s$, $t$ and $u$ channels, whereas the $u$ channel singularity is absent from SYM tree diagrams which are necessarily planar. Likewise, the Jacobians from the scalar integrals do not have singularities in the $u$ channel. This provides us with another way to understand the factor of $u$ that appeared in equation (\ref{eq:gravityproduct}): it cancels a singularity - and prevents a factorization - of the gravity amplitude that is not represented by the particular scalar integral under consideration. (We will see that these $u$ channel factorizations are precisely accounted for by non-planar scalar diagrams.)

For example, consider again the tennis court diagram of figure~\ref{fig:tennis}, now thought of as a sum of Feynman diagrams in ${\cal N}=8$ supergravity. In the corresponding scalar diagram, integrating out the upper right box gives the Jacobian $1/((p+k_3)^2(\ell+k_3)^2)$, exactly as in Yang-Mills. However, using the resummation formula of figure~\ref{fig:sugra}, the contour integral over the Feynman diagrams here gives $(\ell+p)^2\times M_4^{\rm tree}$ and we must take account of the extra $(\ell+p)^2$ when considering factorization limits. In the limit $(\ell+k_3)^2\to0$, triangle diagrams again occur so the scalar integral must not have a residue here. In the limit $(p+k_3)^2\to0$, as before the four-particle tree amplitude factorizes to give Feynman diagrams with the structure of a 2-loop ladder (which we replace by the corresponding scalar integral). Applying the identity $s+t+u=0$ to this tree amplitude shows that in this limit $(\ell+p)^2 \to -(\ell+k_3)^2$. Consequently, in place of equation~(\ref{eq:tennisres}) we instead have
\begin{eqnarray}
{\rm Res}_{(\ell+k_3)^2\to0}\ \frac{C^{\rm tennis}}{(p+k_3)^2(\ell+k_3)^2} &=& 0\\
{\rm Res}_{(p+k_3)^2\to0}\ \frac{C^{\rm tennis}}{(p+k_3)^2(\ell+k_3)^2} &=&
-(\ell+k_3)^2\ C^{2\h{\rm loop}\ {\rm ladder}}\ ,
\end{eqnarray}
so that $C^{\rm tennis} = -\left[(\ell+k_3)^2\right]^2\,C^{2\h{\rm loop}\ {\rm ladder}}$. The $L$-loop ladder coefficients themselves are trivial to find: the calculation proceeds exactly as in SYM, except for the presence of a `$u$-type' factor each time we obtain a four-particle tree amplitude. If the ladder is aligned in the $s$ channel, we avoid triangles by factorizing each intermediate four-particle tree amplitude along its $t$ channel. In this limit, $u \to -s$ for the particular diagram under consideration. But for ladder integrals oriented along the $s$ channel, the $s$ of any subloop is the $s$ of the overall diagram. At the very last step, we do not factorize the final four-particle tree amplitude (because all integrals have been performed), so its $u$ remains. Thus we find
\begin{equation}
C^{L\h{\rm loop}\ {\rm ladder}}=\left(s^Lt\right)\times(-s)^{L-1}u\ M_4^{\rm tree}(k_1,k_2,k_3,k_4)
\label{eq:Lgravityladder}
\end{equation}
in agreement with~\cite{Bern:2007hh}.  Note again that the supergravity coefficients have an extra power of ({\it momentum})$^2$ at each loop compared to the SYM coefficients, in agreement with simple dimensional analysis.

In writing equation~(\ref{eq:Lgravityladder}), we have chosen to separate off the $s^Lt$ factor arising from the Jacobians of the scalar diagrams from the remaining factor arising from the Feynman diagrams. This is to highlight that there is a sense in which the scalar basis of ${\cal N}=8$ supergravity also possesses a hidden dual conformal invariance (at least for the planar integrals): the coefficients of scalar integrals in ${\cal N}=8$ supergravity are given by the same dual conformally invariant expressions as in the SYM case, times the four-particle gravity tree amplitude $M_4^{\rm tree}$, times the various factors of `$u$' of  respective subdiagrams, evaluated in the appropriate factorization limit. Indeed, the striking fact that the basis of scalar diagrams ({\it i.e.} without numerator factors) in ${\cal N}=8$ supergravity is {\it precisely the same} as that in ${\cal N}=4$ SYM (at least up to three loops~\cite{Bern:2007hh}, including non-planar diagrams) makes this `hidden dual conformal invariance' inevitable. The point we wish to make is that this separation may be meaningful: these different factors have a distinct origin. Roughly, one may say that the structure of the loop expansion of ${\cal N}=8$ supergravity has the same conformal invariance property as does ${\cal N}=4$ SYM, while violations of this conformal invariance arise from the classical structure of gravity.

\bigskip

\begin{figure}[t]
\includegraphics[scale=0.45]{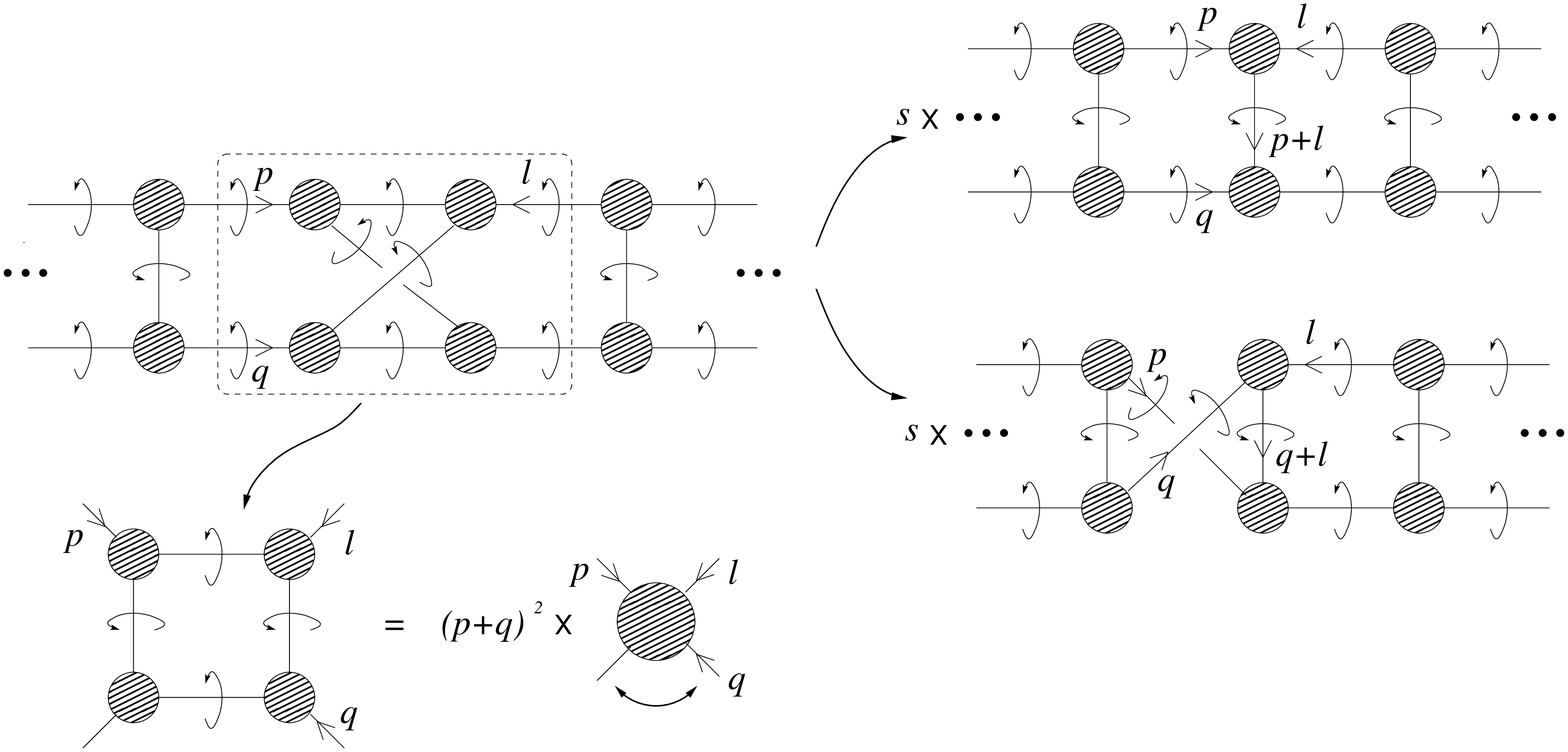}
\caption{A non-planar diagram in ${\cal N}=8$ supergravity.}
\label{fig:nonplanar}
\end{figure}

In gravity, we should not distinguish between planar and non-planar diagrams, so it is important that our technique be able to handle non-planar diagrams also. Consider the set of Feynman diagrams with the structure of a `non-planar ladder' as in figure~\ref{fig:nonplanar}. We first perform the contour integral over the non-planar loop as indicated in the figure. Except for an ordering of the external legs, considered in isolation there is no difference between this subloop and any subloop in a planar ladder. Hence performing the contour integral over the scalar integral gives the Jacobian $1/((p+\ell)^2(q+\ell)^2)$
while the Feynman diagrams yield $(p+q)^2$ times the four-particle tree amplitude. In this case, $(p+q)^2=(k_1+k_2)^2=s$ and remains fixed in any factorization obtainable by choosing a contour for the remaining loops. Because there is no meaningful ordering of the external legs of gravity amplitude, we can sew this four-particle tree amplitude back into the ladder. Considering the two possible factorization limits, we have
\begin{eqnarray}
{\rm Res}_{\ (p+\ell)^2\to0}\
\frac{C_{1234}^{L\h{\rm loop}\,{\rm non}\,{\rm planar}}}{(p+\ell)^2(q+\ell)^2}
&=& s\ C_{1234}^{(L-1)\h{\rm loop}\,{\rm ladder}}\\
{\rm Res}_{\ (q+\ell)^2\to0}\
\frac{C_{1234}^{L\h{\rm loop}\,{\rm non}\,{\rm planar}}}{(p+\ell)^2(q+\ell)^2}
&=& s\ C_{2134}^{(L-1)\h{\rm loop}\,{\rm ladder}}\ .
\end{eqnarray}
As before, in the limits either when $(p+\ell)^2\to0$ or when $(q+\ell)^2\to0$, the other factor becomes $-s$, but now this factor arises from the Jacobians on the left hand side of the previous equations. In these equations, the subscripts on $C$ take account of the ordering of the external legs in figure~\ref{fig:nonplanar}: note that the diagram resulting from factorizing in the $(q+\ell)^2$ channel is not really non-planar, but involves the exchange $k_1\leftrightarrow k_2$. However, it is easy to see that the coefficients $C^{(L-1)\-{\rm loop}\,{\rm ladder}}$ are the same for these two orderings (because $M_4^{\rm tree}$ is invariant under any permutation of the external legs, while under $k_1\leftrightarrow k_2$ we have $t\leftrightarrow u$, each of which appear linearly in equation~(\ref{eq:Lgravityladder})).

In computing the full scattering amplitude, we should sum over all possible permutations of the external legs. Here one must be careful not to overcount: as illustrated in figure~\ref{fig:overcounting}, the three-loop ladder diagram with a non-planar loop is identical after the exchange of $k_1$ and $k_2$. Thus there is a symmetry factor of $1/2$ associated with these non-planar diagrams (which can be thought of as originating from a symmetry factor in the counting of Feynman diagrams). Taking account of this symmetry factor, the coefficient of an $L$-loop ladder with a single non-planar loop is fixed to be
\begin{equation}
C^{L\h{\rm loop}\,{\rm non}\,{\rm planar}}
=-\frac{1}{2}s^2\ C^{(L-1)\h{\rm loop}\,{\rm ladder}} = \frac{1}{2}C^{L\h{\rm loop}\,{\rm ladder}}\ .
\label{eq:nonplanar}
\end{equation}

\bigskip

The same procedure applies no matter how many `twists' are present in the initial ladder, so that the coefficient~(\ref{eq:Lgravityladder}) remains valid for ladders with any degree of non-planarity, so long as appropriate symmetry factors are included. This analysis extends to fix the coefficients of a wide class of non-planar diagrams, both in ${\cal N}=8$ supergravity and ${\cal N}=4$ SYM - with the added subtlety there that the ordering of the particles in a sub-amplitude is significant and must be taken into account in the manipulations analogous to those of figure~\ref{fig:nonplanar}.  There are cases (such as those in figures 2{\it h} \& {\it i} of reference~\cite{Bern:2007hh}) that require new techniques, just as there are non rung-rule diagrams in the planar expansion of ${\cal N}=4$ SYM beyond three loops. It is to these we now turn.

\begin{figure}[t]
\includegraphics[scale=0.45]{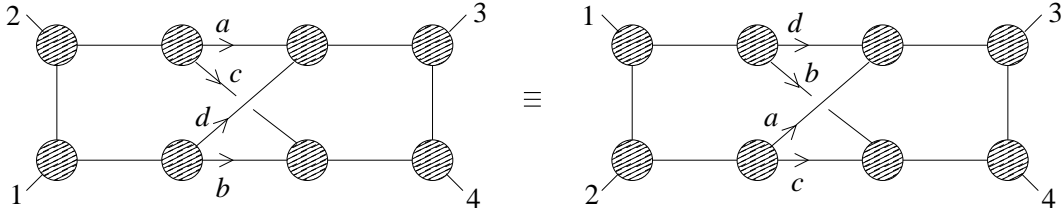}
\caption{Identical non-planar diagrams. To prevent overcounting, these must be included with a symmetry factor of $1/2$ in the computation of the full amplitude.}
\label{fig:overcounting}
\end{figure}

\section{Limitations and Possible Solutions}
\label{sec:lim}

Our discussion is valid for a large class of integrals that
contribute to amplitudes in ${\cal N}=4$ SYM and ${\cal N}=8$ supergravity
but this is clearly not enough. In this section we discuss the two
main limitations of our technique when applied to the planar ${\cal
N}=4$ SYM four-particle amplitude. The reason we restrict to this
case is that it is where we have more to say about the solution to
both problems. Along the way we introduce ideas that might lead to a
derivation of the basis in terms of dual conformally invariant
integrals with their corresponding coefficients.

The basic idea is to use the fact that if it were not for IR
divergencies, conformal invariance in ${\cal N}=4$ SYM would imply that amplitudes must be proportional to the tree amplitude, with a proporationality factor that depends only on conformally invariant combinations of the available kinematic invariants. For amplitude of four (massless) particles, there are no such invariants, so - if we could ignore IR divergencies - the full amplitude would have to be $f(\lambda)A_4^{\rm tree}$ for $f(\lambda)$ some function of the coupling constant. Of course, it is not correct to ignore IR divergencies, but we see that the non-trivial perturbative
expansion must be tied up with the IR singularities. Indeed, we
give evidence that by using the equations coming from the IR
singular behavior, which connect tree-level amplitudes to one-loop
amplitudes, one can derive the rung-rule which connects $L$-loop
amplitudes to $(L+1)$-loop ones.

Integrals in the basis coming from the rung-rule are the first
attempt to reproducing the rung-rule like behavior in Feynman
diagrams. However, starting at four loops, rung-rule integrals develop unphysical singularities that must be canceled; this is also due to the presence of massless particles. These singularites may be removed by introducing further integrals to correct the rung-rule. We give
evidence that the process of removing unphysical singularities is
enough to fix the whole amplitude. This means that, indeed, the
four-particle amplitude is determined to all orders in perturbation theory by IR divergencies.

\subsection{Going Beyond the Four-particle Box: Rung-rule From IR Singularities}
\label{sec:ir}

The main identities (figures~\ref{fig:sym} \&~\ref{fig:sugra}) that allow the computation and physical interpretation of hidden cuts come from the explicit form of
one-loop four-particle amplitudes. However, the same identities could
have been obtained in a different way using the IR singular behavior
of the amplitudes.

The four-particle identity is, in fact, part of a series of
identities valid for $n$ particles that are derived from
the IR singularities of the amplitudes. Before discussing the
equations in more detail, let us point out an interesting point
already for $n=4$. Suppose we are given a piece of an $L$-loop
diagram like the one depicted on the top left of figure
\ref{fig:BasicIR}. As explained in previous sections, we can think
of this as a tree level amplitude in a particular factorization
limit. In figure \ref{fig:BasicIR} we show a tree level amplitude
with a dashed line representing the fact that the corresponding
channel must be suppressed. In this case it is the
$(p+q)^2$-channel. Using the identity that relates the tree
amplitude to a one-loop amplitude on a $T^4$ contour, we find a
connection between the original $L$-loop diagrams and an
$(L+1)$-loop diagram. The dashed line in the $(L+1)$-loop diagram is
there only to encode the information about the factorization channel
that must be used.

\begin{figure}[t]
\includegraphics[scale=0.50]{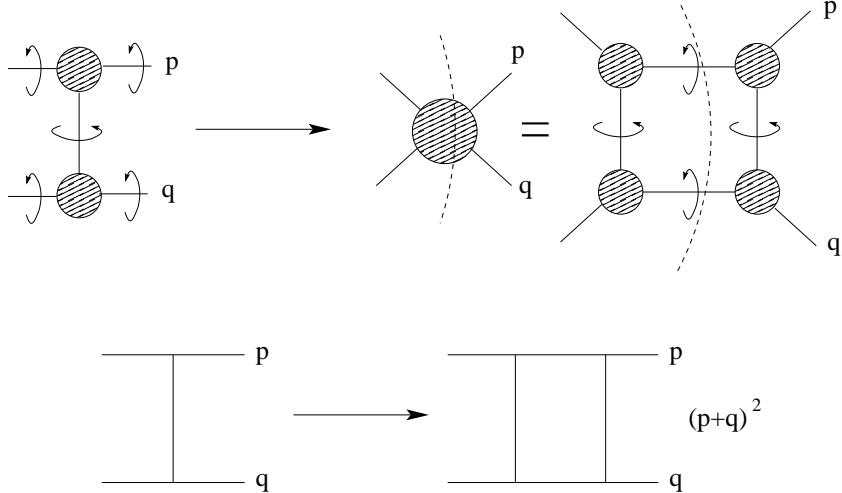}
\caption{Simplest example of the rung rule from IR singular
behavior.}
\label{fig:BasicIR}
\end{figure}

One would like to reproduce this behavior using scalar integrals.
The way to do it is shown on the bottom of figure \ref{fig:BasicIR}.
In this case, the way to encode the information of the dashed line
is by adding a numerator. This is nothing but the rung-rule applied
to the simplest case where one produces a box after adding the rung. The rung-rule can also be used to create polygon subloops that have any number ($\geq 4$) of edges. As noted previously, it is important to understand how to generalize both the techniques of the previous sections and the discussion above to such cases. The key to doing this is in the IR equations which we now discuss in more generality.

The IR singular behavior is well understood in the case when all the
external particles are gluons~\cite{Giele:1991vf,Kunszt:1994mc,Catani:1998bh}. To our knowledge there is no
result in the literature that extends the discussion to all other
particles in the ${\cal N}=4$ SYM multiplet. Here we will simply
assume it is also true for any other states in the ${\cal N}=4$
multiplet. For four and five particles this is guaranteed by SUSY Ward identities; we leave the question of its validity in general as a very important gap in our derivation which deserves further study.

The IR behavior of a one-loop $n$-particle amplitude is
\begin{equation}
\label{eq:ired} \left.A^{(1)}_n\right|_{\rm IR} = A^{\rm
tree}_n\sum_{i=1}^n\frac{1}{\epsilon^2}\left(-\frac{s_{i,i+1}}{\mu^2}\right)^{-\epsilon}
+ {\cal O}(\epsilon^0)\ ,
\end{equation}
where $s_{i,i+1} = (p_i+p_{i+1})^2$ and $\mu$ is an arbitrary mass
scale. Only singularities in consecutive particles can arise because
we are restricting the discussion to leading-color planar partial
amplitudes.

Given that the amplitudes are expressed as a sum over scalar boxes,
which contain singularities in many channels, the
constraint~(\ref{eq:ired}) gives rise to several equations. For
example, the coefficient multiplying the IR singularity in the
$(p_1+p_2+p_3)^2$ channel must vanish. A particularly useful linear
combination of these equations was found by Roiban {\it et. al.}~\cite{Roiban:2004ix}
\begin{equation}
\label{eq:gene} A^{\rm tree}_n = \frac{1}{2}\sum_{j=i+2}^{i-2}
B_{i,i+1,i+2,j+1}
\end{equation}
where indices are understood modulo $n$ and $B_{i,i+1,i+2,j+1}$ is
the coefficient of a box with $K_1 = p_i$, $K_2 = p_{i+1}$, $K_3 =
p_{i+2}+\ldots +p_{j}$ and $K_4 = p_{j+1}+\ldots + p_{i-1}$. Note
that this is true for any choice of $(i)$.

The final ingredient comes from the leading singularity discussed in
section~\ref{sec:1loop}. This gives a representation of the coefficients $B$'s
in terms of products of tree-level amplitudes as follows:
\begin{equation}
\label{eq:sasi} B_{i,i+1,i+2,j+1}  = \sum_{\rm
Multiplet}\int_{\Gamma}d^4\ell\  A^{\rm
tree}(\ell_1,p_i,\ell_2)A^{\rm tree}(\ell_2,p_{i+1},\ell_3)A^{\rm
tree}(\ell_3,{\cal I},\ell_4)A^{\rm tree}(\ell_4,{\cal J},\ell_1)\ ,
\end{equation}
where $\ell_1 = \ell$, $\ell_2 = \ell+p_i$, $\ell_3 =
\ell+p_i+p_{i+1}$, $\ell_4 = \ell +p_i+\ldots +p_j$, ${\cal I} = \{
p_{i+2},\ldots, p_{j} \}$ and ${\cal J}=\{ p_{j+1},\ldots
,p_{i-1}\}$. As in section~\ref{sec:1loop}, the integrand is viewed as a meromorphic function of complex momenta and is integrated over the contour $\Gamma\cong T^4$ defined by $\{
|\ell_i^2|=\epsilon, i=1,2,3,4 \}$. The sum is over all members of the
${\cal N}=4$ multiplet in each internal line. Combining~(\ref{eq:gene}) and~(\ref{eq:sasi}) one finds the equation depicted on the top of figure~\ref{fig:IRrung}.

\begin{figure}
\includegraphics[scale=0.45]{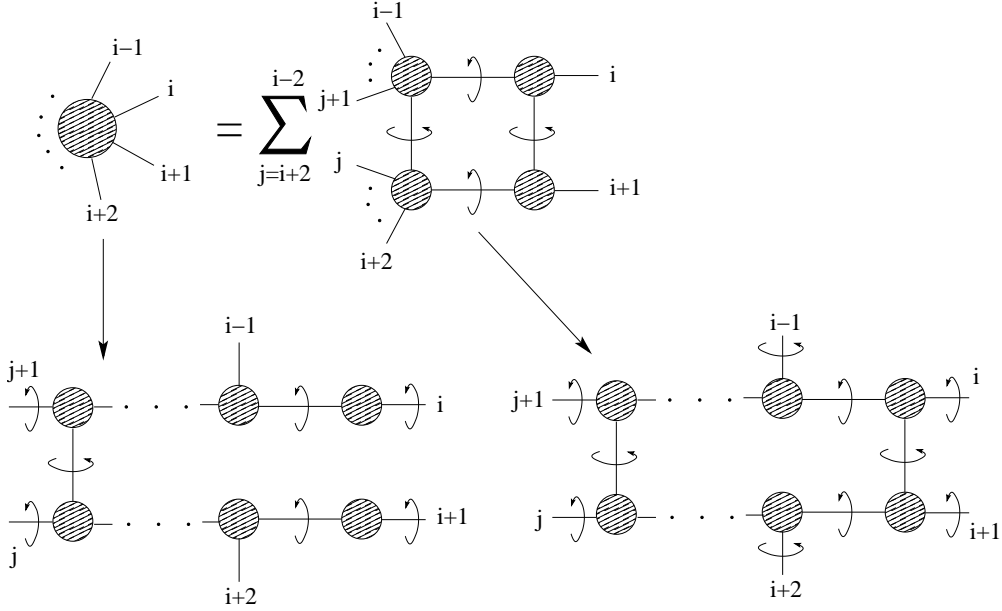}
\caption{Top: Identity obtained from the IR singular behavior.
Bottom: Factorization limits that lead to the rung rule.}
\label{fig:IRrung}
\end{figure}

Now we can see the general pattern that might lead to a derivation
of the rung rule: the IR singular behavior links tree-level
amplitudes to one-loop amplitudes on a special contour of
integration. Embedding this in multiloop integrals would provide a
bridge between $L$- and $(L+1)$-loop integrals.

In order to make the link more precise we consider multiple
factorization limits of $A^{\rm tree}_n$ such that it becomes a tree
with only three-particle amplitudes (for $n=4$ this is achieved by a
single factorization as shown in figure \ref{fig:BasicIR}).
When this tree amplitude occurs as a subdiagram in a multiloop diagram, the `external' momenta of this subdiagram may depend on the momenta running around other loops. The factorization displayed in figure~\ref{fig:IRrung} can then be induced by choosing the full contour of integration to involve going around the singularties where each of the displayed propagators go on-shell. The same kind of factorization must happen on the right hand side of figure~\ref{fig:IRrung} when the `external' momenta are tuned to
produce the factorization on the left. We have indicated the result of such a factorization at the bottom of the figure. It is important that only one of the terms in the sum on the right hand side has the correct factorization, {\it i.e.} that the contour can actually be chosen so as to induce the particular factorization shown in figure~\ref{fig:IRrung}, and we have not been able to prove this in the general case. (Note that when working with complex momenta there are many different ways of achieving the
same factorization of the tree amplitude on the left hand side.) Note that one side is related to the other by addition or removal of a rung. This is indeed the analog for Feynman diagrams of the
rung-rule.

In complete analogy to the four particle case, if we want to model the
behavior of these Feynman diagrams using scalar integrals, one has
to ensure that these integrals have the same value as the Feynman diagrams in all possible factorization channels. In particular, the scalar integral must not have any contribution in factorization channels that are unphysical, or else lead to triangle subdiagrams in the Feynman diagrams. Thus, when we add a rung to the scalar integral as in figure~\ref{fig:IRrung}, we must also introduce a numerator that removes the factorization in the $s_{i,i+1}$ channel. This is exactly the rung rule in it most
general form! In other words, armed with the infra-red relations of figure~\ref{fig:IRrung} we can now apply the technique of the first part of the paper to a wider class of diagrams, including those that do not contain boxes.

\bigskip

It would be interesting to formalize and fill in the gaps of this
argument so that the rung-rule for scalar integrals would have a
purely IR origin. To summarize, the missing steps in the proof are, firstly,
the validity of IR singular equations for any number of external
particles in the ${\cal N}=4$ SYM multiplet and secondly, a more complete and systematic understanding
of the existence and location of the contours necessary to single out the particular factorization on the right hand side of figure~\ref{fig:IRrung}. Note that asking for the IR equations to be valid for arbitrary members of the supermultiplet is weaker than requiring that the Ward identities relate all $n$-particle one-loop amplitudes themselves.

\subsection{Corrections}
\label{sec:corrections}

The fact that rung-rule integrals seem to naturally be related to
the IR singular behavior of the theory leads us to believe that they
are the basic building blocks of the amplitudes. As mentioned in
previous sections, the rung-rule is in fact known to give the full
basis of integrals up to three loops. At four loops and higher the
rung-rule falls short and new integrals must be added. As reviewed in section~\ref{sec:dci}, the missing integrals are supplied by adopting the principle of dual conformal invariance. This provides an
ansatz for the basis of integrals that was used in~\cite{Bern:2007ct} to build a
proposal for the five-loop amplitude which passes many non-trivial
tests. Experience shows that the coefficients of the non rung-rule
integrals always turn out to be $\pm 1$ while those of the rung-rule
ones are always equal to $+1$.

Here we would like to propose the point of view that the scalar
integrals are a representation of the amplitude defined in terms of
Feynman diagrams and that rung-rule integrals give a first
approximation to the amplitude. Starting at four loops one finds
that rung-rule integrals can contain unphysical singularities which
must be removed. Such singularities are removed by adding new integrals into the scalar basis. We call the two new integrals that appear at four loops `first-order corrections' because they correct the rung-rule ones. At five and higher loops, the first-order corrections themselves develop unphysical
singularities which must again be canceled. This is done by adding
second-order corrections. We expect that this process continues
indefinitely and one will find $n^{\rm th}$-order corrections for
any $n$ at sufficiently high loop orders.

The cancelation of spurious singularities fixes the relative sign
between a rung-rule integral and its correction. Given that all
rung-rule integrals come with coefficient one, we find that $n^{\rm th}$
order corrections come with coefficient $(-1)^n$. The requirement that this
assignment of signs be consistent is quite non-trivial and we
comment on it at the end of the section.

\subsubsection{Spurious Singularities}

We start the motivation for corrections by looking at two-mass-easy
(or three-mass) ladder integrals\footnote{Two-mass-easy means that
two diagonally opposed legs are massive ({\it i.e.} represent more
than one external particle) and the other two are massless.}. An
explicit computation of the integral reveals that it has a simple
pole at $st-P^2Q^2=0$, where $s$ and $t$ are the usual channels in
the planar case while $P$ and $Q$ are the total momenta of each of
the diagonally opposed massive legs. Such poles must be spurious
since they do not correspond to physical singularities in a
scattering amplitude. At one loop, this means that the coefficient
of the scalar integral must have a zero at the same location. In
fact, in~\cite{Britto:2004nc} it was shown that the coefficients of
scalar integrals possess a simple zero at $st-P^2Q^2=0$. This fact
was also found for MHV amplitudes in~\cite{Bern:1994zx}.

The presence of unphysical singularities in generic integrals is not
straightforward to detect. A clue which leads to a systematic search
for unphysical singularities comes from the study of ladder
diagrams. Consider the one-loop and two-loop ladders in figure
\ref{fig:OneTwo}. The solid lines represent the numerators that come
from the rung-rule. As explained in section \ref{sec:dci}, it is
convenient to introduce a dual diagram and label points in each of
the faces of the diagram and denote by lines numerator factors which
are given by the sum over momenta crossed by the line all squared.
In the case of the one-loop ladder (or box) the numerator is
$st=x_{24}^2x_{13}^2$. In this case, the vertical solid line
represents $s=x_{24}^2$ while the horizontal one represents
$t=x_{13}^2$. The numerator in the two-loop ladder, represented by
the solid lines, is $st^2=x_{24}^2(x_{13}^2)^2$.

\begin{figure}[t]
\includegraphics[scale=0.50]{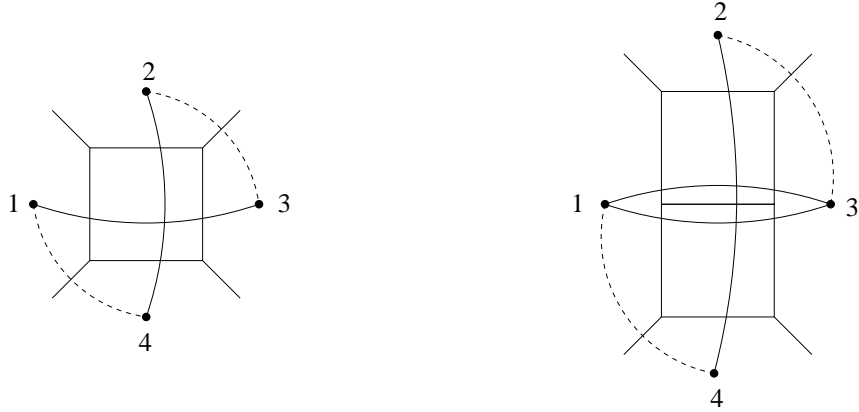}
\caption{Corrections for one- and two-loop three mass or two-mass
easy integrals. We choose to label the momenta on the top right of
the integrals by $P$ and on the bottom left by $Q$. Other momenta
will not enter in the discussion. The solid lines represent the
numerator of the rung-rule diagrams, while the dashed lines
represent the numerators of the required corrections.}
\label{fig:OneTwo}
\end{figure}

Suppose for a moment that the external legs were not on-shell or
that these are subdiagrams of a multiloop integral\footnote{A nice
example is given in figure \ref{fig:FiveCorrection}C taken
from~\cite{Bern:2007ct}. There we can explicitly see the one loop
(together with the numerators) embedded in a larger diagram.}. Now
make one leg on-shell (in the case of an embedded subloop this will
happen in some region of the integration phase space). Then the
integral possesses a pole at $st-P^2Q^2 = x_{24}^2x_{13}^2
-x_{23}^2x_{14}^2$ with the labeling explained in the caption of
figure \ref{fig:OneTwo}. Given that each integral comes with a
factor of $st$ in the numerator, the only piece missing to produce
the desired zero which cancels the unphysical pole is $-P^2Q^2$. We
represent this numerator by the dashed lines in the figure. The
correction is thus obtained by removing the solid lines and
introducing an integral with numerator given by the dashed lines
with a minus sign relative to the integral being corrected. Note
that only the solid lines attached to the dash lines are removed.
For example, in the two loop case, only one of the two horizontal
solid lines must be removed. This means that the new integral has
numerator $-tP^2Q^2$.

\bigskip

This suggests a systematic way to search for singularities: start
with a rung-rule integral, combine numerators (that do not share any
end points) pairwise and join them by dashed lines.

There are two important subtleties to this procedure. Here we will
discuss one which is transparent from the ladder examples and which
will be crucial at higher loops. We postpone the discussion of the
second one to the case where it first appears, {\it i.e.} at four
loops. Suppose as above that the diagrams in figure \ref{fig:OneTwo} are
subdiagrams of a higher-loop amplitude. Then there will be
regions in the higher-loop momentum integration where $P=x_{23}$ and $Q=x_{14}$ are massless while $x_{12}$ and $x_{34}$ are massive. In these regions the dashed lines would
be $x_{12}$ and $x_{34}$ instead of $x_{14}$ and $x_{23}$. Clearly,
these also lead to unphysical singularities that have to be removed.
Therefore the corresponding corrections must also be added.

One might wonder if the new integrals we added also possess
unphysical singularities that must be removed. We will find that
this is indeed the case at five loops and higher. The search for
such singularities is more subtle. The reason is that if we naively
apply the rule we just described to, say, the box with numerator
$x_{23}^2x_{14}^2$ we will find that $x_{12}^2x_{34}^2$ is a
potential singularity. However, from the explicit form of the ladder
we know that $x_{23}^2x_{14}^2-x_{12}^2x_{34}^2=0$ is {\it not} a
singularity!

A simple way to avoid this subtlety is to remember that corrections
come in pairs. In other words, there are always two ways to draw the
dashed lines. Each of them leads to a correction which has a minus
sign relative to the rung-rule integrals (or to the integral they
correct). In computing further corrections any choice of dashed
lines that connects these original two corrections is not a singularity.

Now we are in a position to start discussing corrections to the
rung-rule diagrams of a four-particle amplitude at each loop order.
At one loop, it is clear that all corrections vanish since for four
particles, all external legs are massless. At two loops, the
numerator of the correction is $-tP^2Q^2$. Once again this vanishes
since the external legs are on-shell. Note that any corrections to
the one-loop subdiagrams also vanish here; spurious singularities
are absent again because the external legs are on-shell. One can
easily check that all rung-rule diagrams up to three loops are
correction-free\footnote{Provided one uses a regularization
procedure in which the external legs are kept massless. In the
off-shell regularization procedure we expect the rung rule to be
corrected even at low loop orders.}. We now move to the first case
in which there are non-trivial corrections.

\subsubsection{Corrections at Four Loops}

Applying the rung-rule to construct a basis of integrals at four
loops gives rise to 24 integrals (see figure
\ref{fig:AllFourloop}). They all contribute to the amplitude with
coefficient one.

\begin{figure}
\includegraphics[scale=0.50]{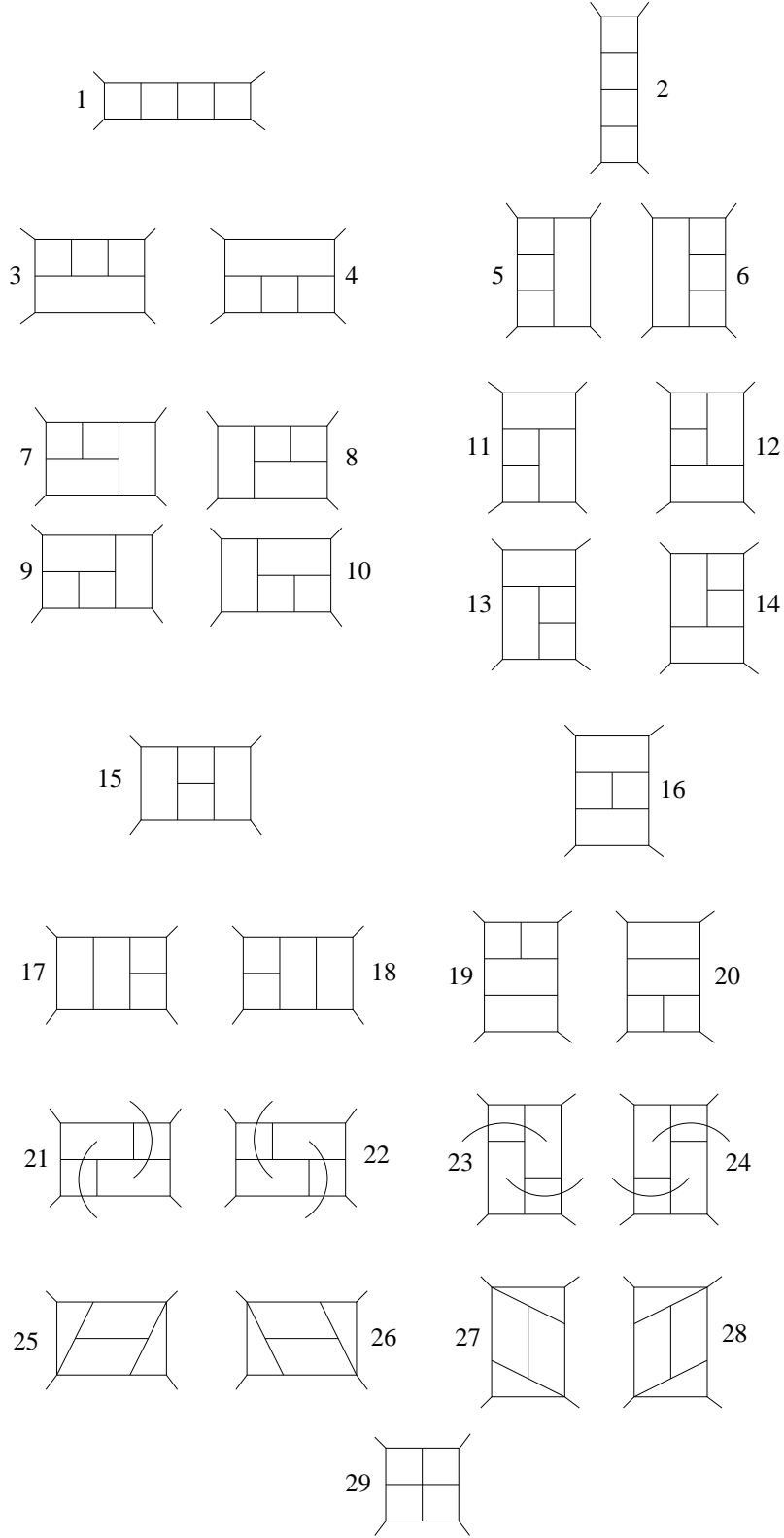}
\caption{Integrals $1-24$ are obtained by applications of the
rung-rule. The numerators are suppressed except in $21-24$ where
they are not uniquely determined. Integrals $25-29$ are
corrections.}
\label{fig:AllFourloop}
\end{figure}

We proceed with the computation of corrections by taking each of the
24 integrals, choosing rung-rule numerators pairwise so that their
corresponding solid lines do not share any end points. For each pair
we draw dashed lines and compute the correction as explained in
detail for the ladder cases above. We find that integrals $1-14$ and $17-20$ contain no unphysical singularities, again because the external legs are on-shell. However, the remaining integrals do contain unphysical singularites and hence need to receive corrections. The results are the following:

\begin{equation}\nonumber
\begin{array}{ccc}
{\rm (Rung-rule)} &\rightarrow& {\rm (Corrections)} \nonumber \\
I_{15} & \rightarrow & \{ I_{25}, I_{26} \} \nonumber \\
I_{16} & \rightarrow & \{ I_{27}, I_{28} \} \nonumber \\
I_{21} & \rightarrow & \{ I_{27}, I_{29} \} \nonumber \\
I_{22} & \rightarrow & \{ I_{28}, I_{29} \} \nonumber \\
I_{23} & \rightarrow & \{ I_{25}, I_{29} \} \nonumber \\
I_{24} & \rightarrow & \{ I_{26}, I_{29} \} \nonumber \\
\end{array}
\end{equation}

Note that a given correction can appear multiple times in the
list. Consider for example $I_{25}$: this corrects both $I_{15}$ and
$I_{23}$. It turns out that the singularities that it corrects are
distinct and happen at different regions of the momentum integration. This
explains the fact that the coefficient of the correction must be
minus one and not minus two: the same integral does not need to be counted twice. In fact, if it was minus two then the zero would not cancel the pole. In figure~\ref{fig:CorrectionFour} we explicitly show how to
compute the correction to one of the two unphysical singularities in
$I_{15}$ and in $I_{22}$.

The fact that to each rung-rule integral in the table one has to add
two corrections should not come as a surprise. It is just a
consequence of the observation made earlier that there are always two ways to choose the
dashed lines. Here we see that if one naively follows that rule for computing corrections
and applies it to, say, $I_{25}$ one would find $I_{26}$ as a
potential correction. However, as discussed earlier the
corresponding singularity is clearly absent.

\begin{figure}[t]
\includegraphics[scale=0.50]{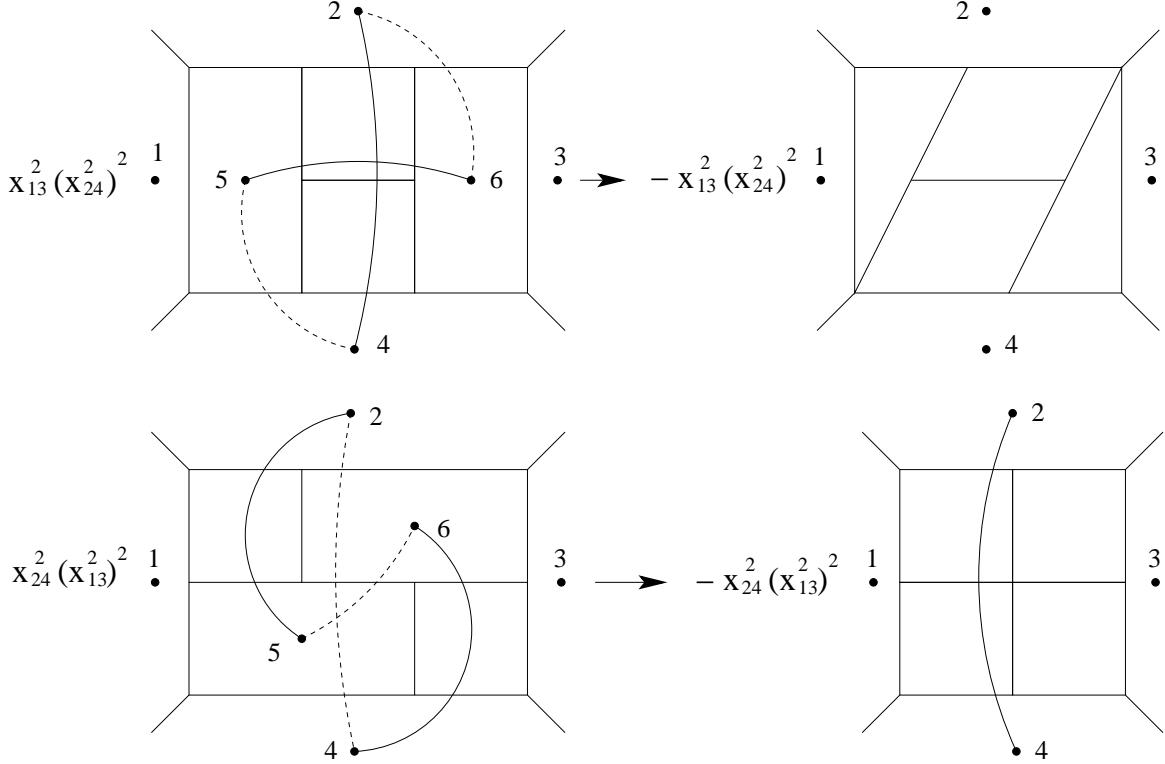}
\caption{Non rung-rule integrals appearing as corrections to cancel the unphysical singularities in $I_{15}$ and in $I_{22}$. Numerator factors relevant to the correction are drawn explicitly; the remaining numerator factors are written in front of the diagrams.}
\label{fig:CorrectionFour}
\end{figure}

\bigskip

There is one more point, or subtlety, which is important to mention
that arises from our pictorial way of computing corrections. Our
procedure should be regarded as a way of finding potential
singularities. In order words, sometimes the procedure of choosing
pairwise rung-rule numerators and joining them by dashed lines leads
to factors of $st-P^2Q^2$ that can never appear as poles in the
rung-rule integral. This means that these naive corrections should
not be added. The only example\footnote{Excluding the trivial ones
discussed earlier.} at four loops comes from integrals $I_{15}$ and
$I_{16}$. In figure \ref{fig:BadCorrection} we show the two
integrals that arise by naively applying the procedure to $I_{15}$.
In this particular case it is simple to understand why the
corresponding singularity is not present in the rung-rule integral:
the only subdiagram that can have a singularity is the two-loop
ladder in the middle of $I_{15}$. None of the two combinations of
$st-P^2Q^2$ shown in the figure can appear in the ladder and hence
are not present in the full diagram.

\begin{figure}[t]
\includegraphics[scale=0.50]{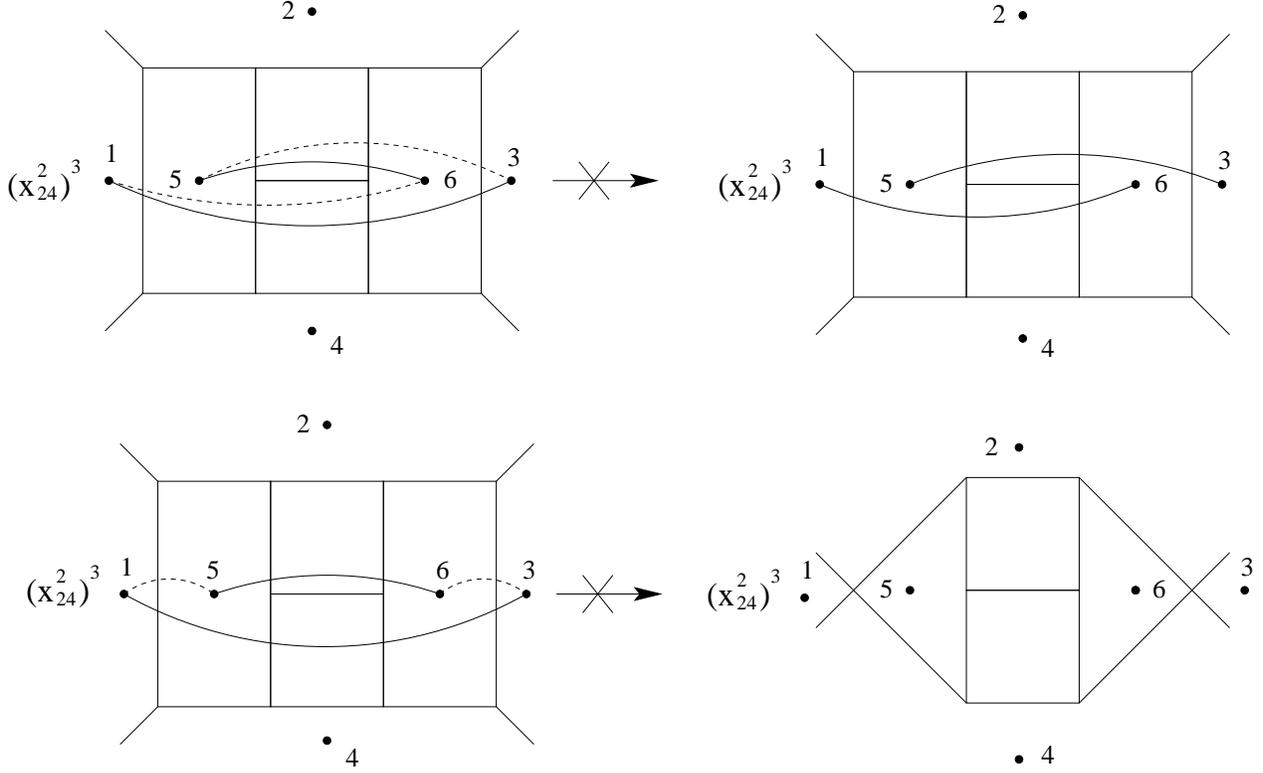}
\caption{Examples of naive corrections that are not needed because
the corresponding singularity is not present. The resulting have dual conformally invariant integrands, but the integrals are not finite.}
\label{fig:BadCorrection}
\end{figure}

At present, we do not have a complete understanding of how to
distinguish true from false singularities in general rung-rule
diagrams. However, it turns out that in all cases where it is
possible to determine that the singularity is absent, the integral
that one naively would have included in the basis violates the
finiteness principle of~\cite{Drummond:2007aua}. It would be very interesting to
establish a connection between the two criteria.

Having finished the process of adding corrections to the rung rule
integrals we should take the new integrals and ask if they need
further corrections. At four loops it is easy to check that no other
corrections are needed.

\subsubsection{Examples at Five-Loops}

At five loops, the whole basis of rung-rule integrals can be found
in~\cite{Bern:2007ct}. In that paper, a proposal for the full
five-loop amplitude was given by using the dual conformally
invariant basis as an ansatz and by determining the coefficients
using various unitarity-based techniques. There it was found that
precisely those integrals that are not finite in the off-shell
regularization come with coefficient zero. The structure of non
rung-rule integrals turns out to be quite complicated. However, all
coefficients turn out to be equal to $\pm 1$. Some of the
coefficients can be understood by relating them to lower-loop
amplitudes but no systematic rule or understanding was given. We
have checked that all the coefficients given in~\cite{Bern:2007ct}
are correctly reproduced by treating non rung-rule integrals as
corrections.

\begin{figure}[t]
\includegraphics[scale=0.50]{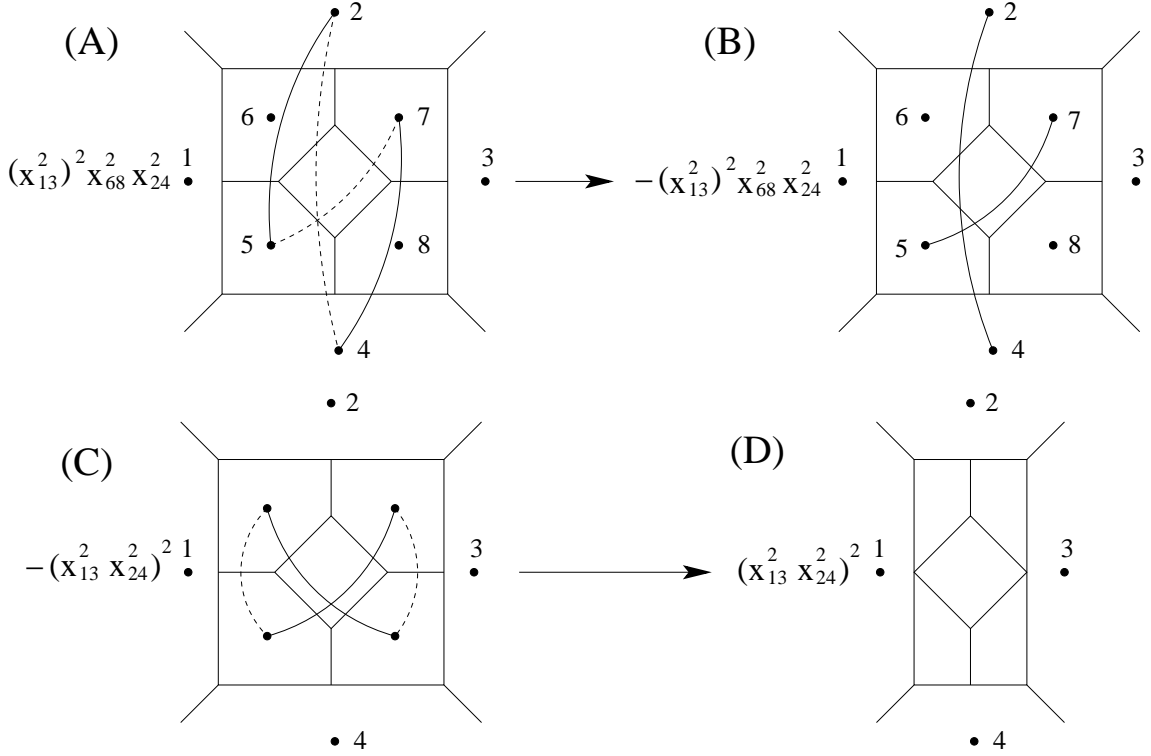}
\caption{A) A rung-rule integral containing an unphysical singularity. B) The first-order correction. C) Re-drawing of B, illustrating that the first-order correction itself contains a further unphysical singularity. D) The second-order correction.} \label{fig:FiveCorrection}
\end{figure}

At four loops, no first-order correction required any further
corrections. At five loops the situation is more interesting and
there are cases when the first-order corrections themselves require
further corrections. A simple example is shown in
figure~\ref{fig:FiveCorrection}. Let us discuss this case in some
detail as all other cases are similar. The rung-rule integral is
shown in figure~\ref{fig:FiveCorrection}A. Solid lines represent the
part of the rung-rule numerator that requires a correction. The
remaining factors of the rung-rule numerator, {\it i.e.}
$(x^2_{13})^2x^2_{68}x^2_{24}$, are written explicitly in front of
the diagram. As before, the first-order corrections are computed by
attaching dashed lines to the solid lines. There are two ways of
doing this and each leads to a correction; here we consider only the
choice of dashed lines explicitly depicted in the figure. 
The resulting non rung-rule integral is obtained by removing the
solid lines while keeping the dashed lines as a new numerator. The
unphysical singularity that is to be removed is located at
$x^2_{25}x^2_{47}-x^2_{57}x^2_{24}=0$. Therefore, the integral in
figure~\ref{fig:FiveCorrection}B must come with a relative minus
sign. Now we consider the first-order correction and ask whether it
can develop further unphysical singularities itself.

Redrawing the same integral in figure~\ref{fig:FiveCorrection}C, but
now representing the numerators $x^2_{68}$ and $x^2_{57}$ by solid
lines, we see that the one-loop subdiagram at the center of the
figure can indeed develop an unphysical singularity (this is exactly
the singularity of the one-loop box discussed earlier). Hence a
further correction is required. The second-order correction is
likewise obtained by removing the solid lines in
figure~\ref{fig:FiveCorrection}C and including the dashed lines as a
new numerator. In this case, these new numerators serve to cancel
some of the propagators in figure~\ref{fig:FiveCorrection}C, and the
resulting integral is shown in figure~\ref{fig:FiveCorrection}D.
Once again, a relative minus sign is needed to cancel the new
spurious singularity of the first-order correction, so the
second-order correction has the same overall sign as the original
rung-rule diagram. It is easy to check that this second-order
correction does not require any further corrections.

Note that there is a second potential singularity manifest in
figure~\ref{fig:FiveCorrection}B, represented by joining $x_2$ to $x_7$ and $x_4$ to $x_5$ by dashed lines. However, this is precisely the other choice of dashed lines in the original
rung-rule integral in figure~\ref{fig:FiveCorrection}A. Therefore,
this is not a true singularity and should not be considered. There are other potential singularities in the first-order correction. However, all of them can be shown to connect pairs of
corrections arising from the same rung-rule and thus are not true
singularities.

\subsubsection{Combining Rules}

Our understanding of the way unphysical singularities arise in
general rung-rule integrals is incomplete at present. However, if we
combine this `correction' point of view with the ansatz that the basis
should involve only dual conformal integrals that are properly
regularized by taking the external legs
off-shell~\cite{Drummond:2007aua}, one can write down a set of rules
that completely specify the scalar integral representing a
four-particle planar amplitude in ${\cal N}=4$ SYM to any loop
order. The rules are as follows:

\begin{enumerate}

\item List all finite dual conformally invariant integrals
at $L$-loops that do not vanish when the external legs are taken to be
massless.

\item Identify all integrals that come from the rung-rule and set
their coefficients to be unity.

\item Compute first-order corrections by taking each rung-rule
integral, combining numerators (that do not share any end points)
pairwise and joining them by dashed lines. Determine if the integral
obtained by removing the solid lines and adding the factors
determined by the dashed lines is in the basis of properly
regularized dual conformally invariant integrals. If so, set its
coefficient to minus one, otherwise discard.

\item Compute second-order corrections by re-applying the
procedure in (3) to the first-order corrections. The integrals in
the basis identified as second-order corrections come with
coefficient plus one. Recall that choices of dashed lines that merely
exchange a given correction to the other correction in the same pair do not
genuinely indicate singularities and therefore should not be considered.

\item Iterate the procedure until all integrals in the basis have
been accounted for. The coefficient of an $n^{th}$ order correction
will be given by $(-1)^n$.

\end{enumerate}

There are two non-trivial consistency conditions that our proposal
has to pass. The first is that all finite dual conformally invariant
integrals must be related to rung-rule ones as corrections of some
order. This is a well defined mathematical problem which might be
within reach. The second - and perhaps more striking - check is that
the assignment of coefficients, {\it i.e.} $\pm1$, of a given
integral has to be consistent. Given that non rung-rule integrals
can be corrections to many different integrals (as we saw in the
examples), this condition is highly non-trivial. In particular, a
given integral must not arise as a $2n^{\rm th}$-order correction to
one rung-rule diagram and as a $(2m+1)^{\rm th}$-order correction to
a different rung-rule diagram: such a situation would lead to a
contradiction. It is thus very important to prove that this can
never happen.

To end this section on corrections, it is easy to prove the converse
of the first consistency condition, {\it i.e} that any correction leads
to a dual conformally invariant integral. In~\cite{Bern:2007ct} it was proven that
the rung-rule procedure only generates dual conformally invariant
integrals. When written in terms of the dual variables, the spurious singularities that need to be removed are always of the form
\begin{equation}
x^2_{ij}x^2_{kl} - x^2_{jl}x^2_{ki} = x^2_{ij}x^2_{kl}\left(
1-\frac{x^2_{jl}x^2_{ki}}{x^2_{ij}x^2_{kl}} \right)\ .
\end{equation}
On the right hand side we have separated out an overall factor of $x^2_{ij}x^2_{kl}$, which we take to be the numerator of the integral being corrected. By assumption this initial integral is dual conformally invariant. The correction is obtained by multiplying by $-x^2_{jl}x^2_{ki}/(x^2_{ij}x^2_{kl})$ which is a conformally invariant cross-ratio, and hence the correction inherits the dual conformally invariance property.

\section{Conclusions and Future Directions}
\label{sec:conclusions}

In this paper we have shown that the study of singularities of Feynman diagrams is a powerful tool to obtain information about the perturbative expansion of scattering amplitudes.

Most of our techniques are especially useful in ${\cal N}=4$ SYM and ${\cal N}=8$ supergravity but we expect them to have applications in theories with less supersymmetry as well. In this first part of the paper, both the resummation identity and the vanishing of sums of Feynman diagrams in certain factorization channels relied on the no-triangle property. Since this is connected to the large amount of supersymmetry\footnote{Even ${\cal N}=1$ SYM scattering amplitudes contain triangles and bubbles (see for example~\cite{Bern:1994cg, Dixon:1996wi}).}, a claim that the techniques may extend to less supersymmetric theories requires some explanation. In the second part of the paper we tried to convey the idea that all the relevant structure is actually determined in terms of the IR singular behavior of the theory. Less supersymmetric (including non-supersymmetric) theories have very well-studied IR singular behavior and we believe it would be very interesting to use the equations coming those singularities to constrain the form of four-particle amplitudes. In the general case triangles and bubbles might be allowed, but this only means that we have to enlarge the basis of scalar integrals (perhaps significantly) in order to agree with the Feynman diagrams in all possible factorization channels as in section~\ref{sec:higherloops}.

\bigskip

We interpreted the rung-rule integrals as a first approximation to the Feynman diagrams, and non rung-rule integrals as `corrections' whose role is to cancel unphysical singularities present in the rung-rule diagrams. Even though our discussion in section~\ref{sec:lim} was restricted to the four-particle planar amplitude in ${\cal N}=4$ SYM, it is reasonable to expect that a similar analysis can be applied to the non-planar part and to ${\cal N}=8$ supergravity. Consider the case of ${\cal N}=8$ supergravity. We have seen that the only differences between the Yang-Mills and gravity computations are the factors of '$u$' present in the gravitational resummation formula. This means that a set of scalar integrals analogous to the rung-rule diagrams of planar ${\cal N}=4$ SYM can easily be identified. Having determined these at a given loop level, one should examine them for unphysical singularities and - if found - complete the basis by adding in new `non rung-rule' integrals as in section~\ref{sec:corrections}.

In a very impressive display of computational power, the full three-loop four-particle integrand of ${\cal N}=8$ supergravity was given in~\cite{Bern:2007hh}. The answer is given in terms of a set of integrals denoted $(a)-(h)$. Following their notation, we would say that integrals $(a)-(g)$ do not receive any corrections while integrals with the topology of $(h)$ and numerator $s^2(\ell_{1,2}^2)^2$ must be corrected. A simpler starting point would perhaps be the three-loop non-planar four-particle amplitude in ${\cal N}=4$ SYM also given in~\cite{Bern:2007hh}. There it is even more transparent which integrals are the basic ones and which are the corrections. Non-planar ${\cal N}=4$ SYM and ${\cal N}=8$ supergravity provide an interesting arena to understand the nature of unphysical singularities as the first non-trivial examples arise show up already at three loops (as we discussed in the text, in the planar case one has to go to four loops).

It is conceivable to expect that by mastering the unphysical singularities and corrections, one would be able significantly simplify the determination of non-planar ${\cal N}=4$ SYM and of ${\cal N}=8$ supergravity integrands at high loop levels.

\bigskip

Returning to the planar four-particle amplitude, a very powerful ansatz for generating the basis of integrals was proposed in~\cite{Drummond:2006rz,Drummond:2007aua}. We reviewed this ansatz, which states that the basis consists of all dual conformally invariant integrals with coefficients $\pm1$. From the IR singular behavior we showed how rung-rule integrals seem to arise naturally. These were shown to satisfy the dual conformal invariance property in~\cite{Bern:2007ct}. In section~\ref{sec:corrections}, we explained that all corrections inherit such a property from their rung-rule progenitor. We hope that by filling in the gaps in our argument, one could obtain a first-principles proof of the finite dual conformal invariance ansatz.

Our construction actually gives more information as it uniquely determines the coefficient in front of each integral. In our approach, the assignment of $\pm1$ coefficients must satisfy highly non-trivial consistency conditions. The simplest of them is that a rung-rule integral should never appear as a correction to another rung-rule integral. It is thus very important to explore these conditions in more detail, at least at six loops where we would expect there to be third-order corrections.

\begin{acknowledgments}

We have benefited from discussions with E. Buchbinder,  X. Liu and
M. Spradlin. The research of FC at Perimeter Institute is supported by the Government of Canada through Industry Canada and by the Province of Ontario through the Ministry of Research \& Innovation. The research of DS is supported by the Province of Ontario through ERA grant \#ER 06-02-293.

\end{acknowledgments}

\end{document}